# A deep learning system for differential diagnosis of skin diseases

Yuan Liu[1], Ayush Jain[1], Clara Eng[1], David H. Way[1], Kang Lee[1], Peggy Bui[1,2], Kimberly Kanada[†], Guilherme de Oliveira Marinho[‡], Jessica Gallegos[1], Sara Gabriele[1], Vishakha Gupta[1], Nalini Singh[1,3,§], Vivek Natarajan[1], Rainer Hofmann-Wellenhof[4], Greg S. Corrado[1], Lily H. Peng[1], Dale R. Webster[1], Dennis Ai[1], Susan Huang[†], Yun Liu[1,*], R. Carter Dunn[1,**], David Coz[1,**]

Affiliations:
[1]Google Health, Palo Alto, CA, USA
[2]University of California, San Francisco, CA, USA
[3]Massachusetts Institute of Technology, Cambridge, MA, USA
[4]Medical University of Graz, Graz, Austria

[†]Work done at Google Health via Advanced Clinical.
[‡]Work done at Google Health via Adecco Staffing.
[§]Work done at Google Health.
*Corresponding author: liuyun@google.com
**These authors contributed equally to this work.

**Abstract**

Skin and subcutaneous conditions affect an estimated 1.9 billion people at any given time and remain the fourth leading cause of non-fatal disease burden worldwide. Access to dermatology care is limited due to a shortage of dermatologists, causing long wait times and leading patients to seek dermatologic care from general practitioners. However, the diagnostic accuracy of general practitioners has been reported to be only 0.24-0.70 (compared to 0.77-0.96 for dermatologists), resulting in over- and under-referrals, delays in care, and errors in diagnosis and treatment. In this paper, we developed a deep learning system (DLS) to provide a differential diagnosis of skin conditions for clinical cases (skin photographs and associated medical histories). The DLS distinguishes between 26 of the most common skin conditions, representing roughly 80% of the volume of skin conditions seen in a primary care setting. The DLS was developed and validated using de-identified cases from a teledermatology practice serving 17 clinical sites via a temporal split: the first 14,021 cases for development and the last 3,756 cases for validation. On the validation set, where a panel of three board-certified dermatologists defined the reference standard for every case, the DLS achieved 0.71 and 0.93 top-1 and top-3 accuracies respectively, indicating the fraction of cases where the DLS's top diagnosis and top 3 diagnoses contains the correct diagnosis. For a stratified random subset of the validation set (n=963 cases), 18 clinicians (of three different training levels) reviewed the cases for comparison. On this subset, the DLS achieved a 0.67 top-1 accuracy, non-inferior to board-certified dermatologists (0.63, p<0.001), and higher than primary care physicians (PCPs, 0.45) and nurse practitioners (NPs, 0.41). The top-3 accuracy showed a similar trend: 0.90 DLS, 0.75 dermatologists, 0.60 PCPs, and 0.55 NPs. These results highlight the potential of the DLS to augment the ability of general practitioners who did not have additional specialty training to accurately diagnose skin conditions by suggesting differential diagnoses that may not have been considered. Future work will be needed to prospectively assess the clinical impact of using this tool in actual clinical workflows.

**Introduction**

Skin disease is the fourth leading cause of nonfatal disease burden globally, affecting 30-70% of individuals and prevalent in all geographies and age groups[1]. Skin disease is also one of the most common chief complaints in primary care, with 8-36% of patients presenting with at least one skin complaint[2,3]. However, dermatologists are consistently in short supply, particularly in rural areas, and consultation costs are rising[4,5]. Thus, the burden of triage and diagnosis commonly falls on non-specialists such as primary care physicians (PCPs), nurse practitioners (NPs), and physician assistants[6–8]. Because of limited knowledge and training in a specialty with hundreds of conditions[9], diagnostic accuracy of non-specialists is only 24-70%[10–13], despite the availability and use of references such as dermatology textbooks, UpToDate[14], and online image search engines[15]. Low diagnostic accuracy can lead to poor patient outcomes such as delayed or improper treatment.

To expand access to specialists and improve diagnostic accuracy, store-and-forward teledermatology has become more popular, with the number of programs increasing by 48% in U.S. non-governmental programs between 2011 and 2016[16]. In store-and-forward teledermatology, digital images of affected skin areas, typically captured with digital cameras or smartphones, are transmitted along with other medical information to a dermatologist. The dermatologist then remotely reviews the case and provides consultation on the diagnosis, work-up, treatment, and recommendations for follow-up. This approach has been shown to result in similar clinical outcomes compared to conventional consultation in dermatology clinics[17], and improved satisfaction from both patients and providers[18].

The use of artificial intelligence tools may be another promising method of broadening the availability of dermatology expertise. Recent advances in deep learning have facilitated the development of artificial intelligence tools to assist in diagnosing skin disorders from images. Many prior works have focused on the visual recognition of skin lesions from dermoscopic images[19–26], which require a dermatoscope. However, dermatoscopes are usually inaccessible outside of dermatology clinics and are unnecessary for many common skin diseases. By contrast, others have attended to clinical photographs. For example, Esteva et al. applied deep learning to photographs of skin cancers to distinguish malignant from benign variants[27]. Han et al. developed a region-based classifier to identify onychomycosis in clinical images[28]. Yang et al.[29] presented a new visual representation to diagnose up to 198 skin lesions using a dataset of 6,584 clinical images from an educational website[30,31]. Some of these works also reported comparable performance to experts on binary classification tasks (benign vs. malignant) or on skin lesion conditions[22–24,27]. Though the majority of the papers examined individual skin lesions, dermatologic conditions seen in routine practice more commonly include non-cancerous conditions such as inflammatory dermatoses and pigmentary issues[32]. These skin problems have yet to be addressed despite their high prevalence and similarly low diagnostic accuracy by non-specialists[19–21,27–30,33,34]. Moreover, prior work has focused on predicting a single diagnosis, instead of a full differential diagnosis. A differential diagnosis is a ranked list of diagnoses that is used to plan treatments in the common setting of diagnostic ambiguity in dermatology, and can capture a more comprehensive assessment of a clinical case than a single diagnosis[35].

     In this paper, we developed a deep learning system (DLS) to identify 26 of the most common skin conditions in adult cases that were referred for teledermatology consultation. Our DLS provides several advances relative to prior work. First, instead of a single classification between a small number of conditions, our DLS provides a differential diagnosis across 26 conditions that include various dermatitides, dermatoses, pigmentary conditions, alopecia, and lesions, to aid clinical decision making. Second, instead of relying only on images, our DLS leverages additional data that are available to dermatologists in a teledermatology service, such as demographic information and the medical history. Third, the DLS supports a variable number of input images, and the benefit of using multiple images was assessed. Finally, to understand the potential value of the DLS, we compared the DLS's diagnostic accuracy with board-certified clinicians of three different levels of training: dermatologists, PCPs, and NPs.

# RESULTS

## Overview of approach

Our DLS has two major components: a variable number of deep convolutional neural network modules to process a flexible number of input images, and a shallow module to process metadata which includes patient demographic information and medical history (Fig. 1 and Supplementary Table 1). To develop and validate our DLS, we applied a temporal split to data from a teledermatology service: the first approximately 80% of the cases (years 2010-2017) were used for development, while the last 20% (years 2017-2018) were used for validation (Table 1). This simulates a "prospective" setting where the model is developed on past data and validated on data collected "in the future", and is arguably a form of external validation[36]. To avoid bias, we ensured that no patient was present in both the development and validation sets. Each case in the development set was then reviewed by a rotating panel of 1 to 29 dermatologists to determine the reference standard differential diagnosis, while each case in the validation set was reviewed by a panel of three U.S. board-certified dermatologists (Methods). After excluding cases with multiple skin conditions and those that were non-diagnosable, 14,021 cases with 56,134 images were used for development, while 3,756 cases with 14,883 images were used for validation (validation set "A"; a smaller subset "B" was used for comparison with clinicians and is described in the relevant section). In total, 53,581 dermatologist reviews were collected for development and 11,268 reviews for validation.

## DLS performance

The DLS's top differential diagnosis in validation set A had a "top-1 accuracy" (accuracy across all cases) of 0.71 and an average "top-1 sensitivity" (sensitivity computed for each condition and averaged) across the 26 conditions of 0.60 (Fig. 2a and Extended Data Table 1). When the DLS was allowed three diagnoses (for example to mimic a clinical decision support tool that suggests a few possibilities for the clinician's consideration), the DLS's top-3 accuracy rose to 0.93 and average top-3 sensitivity across the 26 conditions rose to 0.83. To ensure that the DLS was not biased against different skin tones, we evaluated DLS accuracy stratified by Fitzpatrick skin type (Extended Data Table 2). Among the Fitzpatrick skin types that comprised at least 5% of the data (types II-IV), the top-1 accuracy ranged from 0.69 to 0.72, and the top-3 accuracy ranged from 0.91 to 0.94. Additional subanalyses based on self-reported demographic information (i.e., age, sex, race and ethnicity) are also presented in Extended Data Table 2. Evaluation of the DLS's overall differential diagnosis using the average overlap metric[37,38] yielded 0.67 overall (Fig. 2b), and 0.66-0.68 when stratified by Fitzpatrick skin types (Extended Data Table 2). The DLS performance across the 26 conditions is presented in Extended Data Fig. 1a.

### DLS performance compared with clinicians

To compare DLS performance with clinicians, validation set A was randomly subsampled using stratified sampling by condition. This resulted in 963 cases with 3,707 images ("validation set B") that was relatively enriched for the rarer conditions (e.g., 2-5% prevalance in "B" compared to below 1% in "A"). Eighteen clinicians of three different levels of training (dermatologists, PCPs, and NPs, all of whom were board-certified) graded validation set B. On this smaller dataset, the DLS achieved a top-1 accuracy of 0.67, compared to 0.63 for dermatologists, 0.45 for PCPs, and 0.41 for NPs (Fig. 2a). The DLS was non-inferior to the dermatologists at a 5% margin (p<0.001). The top-3 accuracy was substantially higher at 0.90 for the DLS, compared to 0.75 for dermatologists, 0.60 for PCPs, and 0.55 for NPs. Consistent with the top-1 and top-3 accuracies, evaluation of the full differential diagnosis using the average overlap metric yielded 0.63 for the DLS, compared with 0.58 for dermatologists, 0.46 for PCPs, and 0.43 for NPs. The average top-1 and top-3 sensitivities across the 26 conditions followed the same trend (Extended Data Fig. 1b and Extended Data Table 1). Representative examples of cases that were missed by one or more PCPs or NPs are shown in Fig. 3a-e.

### Subgroup analysis

Next, we assessed the DLS's ability to distinguish between conditions that present similarly and can be misidentified in clinical settings, and compared the DLS to clinicians as before (see the "Conditions in the subcategory" column of Table 2 for definitions of the subgroups). The first analysis distinguished between malignant vs. benign growths. Note that in this and subsequent subanalyses, the DLS and clinicians could have determined the case belonged to neither category (e.g. neither a malignant nor a benign growth; i.e. not a growth at all). Because the decision to biopsy depends on whether malignant conditions are in the differential, in this "growths" subgroup analysis, we focused on the top-3 sensitivity for malignant growths. The DLS's top-3 sensitivity of 0.88 was comparable with that of dermatologists (0.89), and higher than that of both PCPs and NPs (0.69 and 0.72, respectively).

The second subgroup analysis distinguished between infectious vs. noninfectious cases of erythematosquamous and papulosquamous skin diseases, the DLS was more sensitive than the clinicians at identifying the infectious subcategory (top-1 sensitivity = 0.75, compared to clinicians' range of 0.48-0.68; top-3 sensitivity = 0.91, compared to clinicians' range of 0.60-0.85). The DLS was also more sensitive at identifying the non-infectious subcategory (top-1 sensitivity = 0.67, compared to clinicians' range of 0.43-0.49; top-3 sensitivity = 0.95, compared to clinicians' range of 0.55-0.62).

The last subgroup deals with two types of hair loss: alopecia areata and androgenetic alopecia. The sensitivity of the DLS for alopecia areata (top-1 sensitivity = 0.77, top-3 sensitivity = 0.86) was higher than PCPs and NPs (0.45-0.59 and 0.64-0.77 for top-1 and top-3 respectively), but not higher than dermatologists (top-1 sensitivity = 0.80, top-3 sensitivity = 0.91). For androgenetic alopecia, the DLS had a top-1 sensitivity of 0.79 and a top-3 sensitivity of 0.91, which was higher than the dermatologists at 0.69 and 0.84, and substantially higher than PCPs and NPs (0.37-0.43 and 0.22-0.29, respectively).

Importance of input data: images versus demographics and medical history

We examined the importance of each of the different input data to the DLS. Among the 45 types of non-image metadata (demographic information and medical history, detailed in Supplementary Table 1), the type of self-reported skin problem (e.g. "acne", "hair loss", or "rash"), history of psoriasis, and the duration of the chief complaint (skin problem) had the greatest impact on accuracy (Fig. 4a).

For image inputs, DLS's performance dramatically improved when more than one image was provided, and plateaued when there were at least five images (Fig. 4b, blue line). This trend was preserved when the non-image metadata were also withheld from the DLS (Fig. 4b, red line). Compared to withholding metadata from the DLS that was developed in the presence of metadata, training another DLS that uses only images (so that it does not "rely" on metadata) yielded a small improvement (Fig. 4b, green line). Finally, saliency analysis via integrated gradients[39] highlighted those regions of the image where a skin condition was visible, suggesting the DLS had generally learned to focus on the right region of interest when making the prediction (Fig. 3a-e).

We also examined the effect of training dataset size on the performance of the DLS, and observed that more training data led to a better top-1 accuracy, though with diminishing return after 10,000 cases (Supplementary Fig. 7).

## DISCUSSION

In this study, we developed and validated a DLS to identify 26 of the most common skin conditions that were referred for a teledermatology consult, representing roughly 80% of cases that present in a primary care setting[1,32,40–42]. Among these cases, the DLS's top-1 diagnostic accuracy was non-inferior to dermatologists and higher than PCPs and NPs. Moreover, the DLS's high top-3 accuracy and average overlap metric suggest that the DLS's full differential diagnosis is relatively complete, and may help alert clinicians to differential diagnoses that they may not have considered.

Providing assistance with a differential diagnosis instead of predicting a single diagnosis is particularly important in dermatology. Because most skin conditions are not verified with pathology, the differential diagnosis is used for decision making around workup and treatment. If all conditions in the differential diagnosis share the same treatment, a single diagnosis may not be clinically necessary as the clinician can proceed with treatment. For example, if the differential diagnosis includes eczema and psoriasis, the clinician may choose to start treatment with topical steroids without having a single diagnosis. If the diagnoses on the differential have opposing treatments (e.g. treatment for one condition on the differential may aggravate another diagnosis on the differential), a clinician can still consider this group of diagnoses together to determine a workup or initiate treatment. For example, if a differential diagnosis included both tinea and eczema, the clinician might perform an in-office KOH exam. If this exam is not possible, the clinician may monitor responsiveness to empiric treatment with a topical antifungal, which could treat tinea and likely not worsen eczema. By contrast, a topical steroid could worsen the condition if it was actually tinea. Another example is a differential including both melanoma and benign nevus. The presence of melanoma on a differential, even if not the most suspected diagnosis, may prompt a clinician to biopsy the lesion to rule out this dangerous clinical entity. In all these situations, the DLS may be an effective aid to non-specialist clinicians by helping them arrive at both a primary diagnosis and a more complete differential diagnosis. Dermatologists in "store-and-forward" teledermatology (where dermatologists review cases asynchronously) could also potentially use such a DLS to help rapidly triage cases.

      To better understand the specific impact of the DLS in challenging diagnostic situations, our subgroup analyses examined several conditions that have similarities in visual presentation, and where the diagnostic accuracy between conditions in different groups can affect the appropriateness of the subsequent clinical decisions. The three subgroup analyses were: individual growths, erythematosquamous and papulosquamous skin disease, and hair loss. For individual growths, malignant lesions should have subsequent biopsy or excision whereas patients with benign lesions can be reassured. The top-3 sensitivity for malignant lesions is important because the inclusion of a diagnosis of malignancy on the differential diagnosis may prompt a clinician to

obtain a specimen for pathology even if it is not the primary suspected diagnosis. For erythematosquamous and papulosquamous skin disease, these eruptions can be clinically similar with erythema and scaling though they can have very different etiologies and treatment plans. High sensitivity is particularly important as first-line treatment of the non-infectious entities is often with a topical steroid, which conversely would make an infectious process like tinea more resistant to treatment and can even hinder diagnosis (e.g. tinea incognito[43]) at future appointments . Additionally, the inclusion of tinea as a potential diagnosis can prompt a clinician to do a KOH exam for confirmation. For hair loss, the two conditions have different etiologies, possible work-up and treatment options.  Distinguishing one from the other could allow a clinician to start first-line therapies and possible workup for these conditions. In the first subgroup, the DLS was very "specific"; i.e. it was able to correctly identify the "negative" subcategory of benign growths. Despite a lower top-1 sensitivity for malignant lesions, the DLS had a high top-3 sensitivity which is on-par with dermatologists. For erythematosquamous and papulosquamous skin disease, and hair loss, the DLS was accurate at detecting both subcategories in each subgroup. Overall, the DLS had substantially higher sensitivities than non-specialty clinicians (with deltas ranging from 2% to 57% for top-1, and 9% to 54% for top-3) in these subgroups. This suggests that the DLS may be particularly valuable in helping determine the workup or initiate treatment based on a working diagnosis.

Overall in this study, dermatologists were substantially more accurate than PCPs and NPs. These results were not surprising as the majority of the cases were sent by primary care providers to a teledermatology service, and presumably the clinician had found diagnostic difficulty for a significant proportion of these cases. Though not strictly comparable because of differences in study design, the low accuracies we observed (36-50%, Supplementary Table 2) are in line with those previously reported (24-70%). These numbers serve to highlight the challenging nature of this classification task that incorporates both visual cues and non-visual information, and underscores the need for decision support tools for non-specialists.

Two conditions in particular seemed challenging based on low clinician accuracies (Extended Data Fig. 1b): allergic contact dermatitis (ACD) and post-inflammatory hyperpigmentation (PIH). Similarly, agreement between dermatologists defining the reference standard was relatively low (Supplementary Table 3). To understand these conditions better, two dermatologists not involved in the reference standard or as comparator clinicians reviewed the ACD and PIH cases. 8 out of the 27 ACD cases were found to be clinically difficult because they did not have a "classic" visual presentation, thus causing ambiguity. Though the clinicians were asked to use the most specific term possible, some of the comparator clinicians used the more general label of "contact dermatitis". However, contact dermatitis also encompasses irritant contact dermatitis, which has a different etiology, workup, counseling, and treatment interventions, a concept that may not be familiar to non-dermatologist graders. This lack of specificity had prompted us to (a priori) categorize contact dermatitis under "Other", which was deemed to be an incorrect answer for ACD cases because we did not consider "partial correctness" in our analysis. On the other hand, PIH was commonly "misdiagnosed" by the comparator clinicians as they often attempted to label what the primary process leading to the PIH could have been. To ensure that that these complexities did not cause our DLS evaluations to be overly optimistic, we recomputed the sensitivities excluding these two conditions for the DLS and all clinicians (Supplementary Table 4), finding 2-4% improved performance for both the DLS and clinicians, with no change in conclusions. More generally, the same conclusions also applied for subanalysis based only on easier cases (where two or three of the reference standard dermatologists agreed on the primary diagnosis), with better performance for both the DLS and clinicians (Supplementary Fig. 1-2).

Previous studies in this area generally have not focused on providing diagnostic assistance in a more generalized workflow, but instead have been focused on early screening of skin cancer, and thus limited to a narrower scope of conditions (e.g. melanoma or not), or on more standardized images that require specialized equipment (i.e., dermoscopic images). Of the studies that attempt to tackle a broader range of conditions[29,30,34], the datasets were either often educational in nature, leading to potential bias towards cases with more typical presentations or unusually severe cases that prompted pathologic confirmation[29,30], or a simplification of labels towards a mix of morphological descriptions (e.g. erythema/"redness") and diagnoses that are too broad to guide clinical workup or treatment (e.g. hair loss without further details)[34]. As a result, the utility of these works in actual clinical settings are unclear. By contrast, the images in our data were taken by different medical assistants across 17 sites, representing a wide variety of lighting conditions, perspective, and backgrounds. Our dataset is also representative of cases that required dermatology consults, and the conditions which our DLS predicts are specific enough to guide a clinician to next steps in clinical care. However, due to the impracticality of performing exhaustive tests or biopsies for all skin conditions, there exists inevitable diagnostic uncertainty in actual clinical settings. To help resolve this, our DLS learns to predict a differential diagnosis instead of a single diagnosis, enabling a decision support tool that surfaces potential diagnoses for clinicians to consider.

Our DLS can potentially augment the current clinical workflow in a primary care setting in several ways. First, the DLS can prompt clinicians to include on their differential a diagnosis that they would not have previously considered. The DLS may thus prevent misdiagnosis, delay to care, and improper treatment which can lead to poor clinical outcomes, a bad patient experience, and increased costs of care. Second, by helping to improve the accuracy of non-dermatologists, the DLS may enable dermatologists to focus on cases that are further along in the care process or which require specialized dermatologic care. Finally, the DLS can aid in the referral triage process. With challenges to access, it is important to identify referred cases as urgent versus non-urgent. If the non-dermatologist clinician provides a more accurate diagnostic assessment of the patient at the time of referral, the patient can be more appropriately triaged for an appointment.

On the technical aspect, while most prior work used only a single image as input, our DLS integrates information from both metadata and one or more images. We further quantify the magnitude of improvement as metadata or more images are provided for each case. Similarly, dermatologists in a teledermatology setting look at multiple images to better appreciate the three-dimensional and textural aspects of the skin findings. We also show that visual features alone enable reasonable diagnostic accuracy by the DLS, and accuracy improves with more images, albeit with diminishing returns after 2-3 images. This has implications on the number of images required for broader real-world use: a single image is likely suboptimal but more than five provides marginal benefit. The addition of metadata such as demographic information and medical history provides a 4-5% consistent improvement independent of the number of images available, with most of the benefit coming from a handful of features out of the 45 provided. This suggests that a few simple questions may be sufficient to capture most of the diagnostic accuracy benefits. Moreover, even the most "important" metadata, i.e., the type of self-reported skin problem, will cause an average of 1.2% reduction in top-1 accuracy when its value is likely incorrect. This suggests that our DLS is relatively robust to metadata error.

Our study has several limitations. First, we did not have a completely external dataset for validation, but instead adopted a prospective-like design by splitting the data temporally. This mimics developing a DLS using several years of retrospective data at a teledermatology practice (which served 17 sites across 2 states), and then validating that DLS prospectively at the same practice on data collected over the next year. To aid generalization beyond the specific metadata available in this dataset, we also have trained a version of the DLS that uses only images as input (Fig. 4b), which may be more easily applicable to practices without or with different metadata. Second, our data did not have pathologic confirmation. Instead, our reference standard for each case was based on aggregating the differential diagnoses of a panel of board-certified dermatologists ("collective intelligence", see Methods and Supplementary Methods for in-depth analysis). Ambiguities in diagnosis do exist in clinical practice, which makes it challenging to evaluate the accuracy of clinicians and DLS, especially for conditions like rashes which are not typically biopsied. Thirdly, as our dataset was de-identified, only structured metadata were available to both the DLS and the clinicians. While useful, it is less rich than free text clinical notes or an in-person examination. Though we were unable to assess this directly, the lack of more comprehensive information may have lowered the diagnostic accuracy of both clinicians and the DLS. With regards to the top-3 metrics, though instructed to, the clinicians provided fewer than 3 diagnoses when sufficiently confident in their first few diagnoses. Thus the clinicians may have higher top-3 metrics if forced to provide at least 3 diagnoses. Lastly, actual clinical cases may present with multiple conditions at the same time. In principle, multiple conditions may be handled as several single-condition diagnoses, though treatment plans may be more complex. In this study, however, multiple conditions were used as an exclusion criteria (Table 1). Future work will also need to assess the generalizability of the DLS to data from additional sites spanning more countries and states, and cases imaged on a greater variety of devices (see Methods).

To conclude, we have developed a DLS to identify 26 of the most common skin conditions at a level comparable to board-certified dermatologists, and more accurate than general practitioners. Our approach could be directly applied to store-and-forward teledermatology by assisting clinicians in triaging cases, thus shortening wait times for specialty care and reducing morbidity that results from skin diseases. Within (in-person) primary care, our algorithm could help improve the accuracy of non-dermatologists, thus allowing the treatment to be initiated instead of waiting for referrals.

## METHODS

### Dataset

The dataset for this study consisted of adult cases from a teledermatology service serving 17 primary care and specialist sites from 2 states in the U.S.. Cases were predominantly referred by medical doctors, doctors of osteopathic medicine, NPs, and physician assistants. Each case contained 1-6 clinical photographs of the affected skin areas taken by medical assistants or trained nurses (approximately 75% of cases had six or fewer images; for cases with more images, six images were randomly selected) and metadata such as patient demographic information and medical history (for a complete list, see Supplementary Table 1). Images were taken on a mix of devices: Canon point-and-shoot cameras and Apple iPad Minis. All images and metadata were de-identified according to Health Insurance Portability and Accountability Act (HIPAA) Safe Harbor prior to transfer to study investigators. The protocol was reviewed by Advarra IRB (Columbia, MD), which determined that it was exempt from further review under 45 CFR 46.

To mimic a prospective design, the dataset was split in a 80:20 ratio based on the submission date of the case: the development set contained cases from 2010-2017, while the validation set contained cases from 2017-2018 (Table 1). The validation set was filtered to ensure no patient overlap with the training set and thus prevent any potential label leakage due to the presence of cases from previous visits in the training set. This validation set "A" was further subsampled to reduce class imbalance among the skin conditions of interest to obtain validation set B (Table 1). Selection of skin conditions are described in a subsequent section ("Labeling tool and skin condition mapping").

Reference Standard Labeling: Validation Set

Because of the impracticality of pathologic confirmation of all diagnoses (e.g. rashes are rarely biopsied), each case's differential diagnosis for the validation set was provided by a rotating panel of three dermatologists from a pool of 14 U.S. board-certified dermatologists. The dermatologists had 5-30 years of experience (average 9.1 years, median 6.5 years), and were actively seeing patients in clinic. The dermatologists also passed a certification test on a small number of cases to ensure that they were comfortable with grading cases using the labeling tool (Supplementary Table 5 and Supplementary Fig. 3). Every dermatologist graded each case (clinical photographs, demographic information, and medical history) independently for the presence of multiple skin conditions, diagnosability (e.g., due to poor image quality, minimal visible pathology, or limited field-of-view), and up to three differential diagnoses using a custom annotation tool (see "Labeling tool and skin condition mapping"). Cases labeled as containing multiple skin conditions or as undiagnosable by the majority of the dermatologists were excluded from the study.

Because grades from individual graders can demonstrate substantial variability, to determine the reference standard, we aggregated the differential diagnoses of the three dermatologists that reviewed each case based on a previously proposed "voting" procedure[44] (see Supplementary Methods and for details and Supplementary Fig. 8 for an example). Briefly, for each grader, each diagnosis was first mapped to one of 421 conditions (see "Labeling tool and skin condition mapping" below), and duplicate mapped conditions were removed. "Votes" for each of these mapped conditions were summed across the three dermatologists based on the relative position of each diagnosis within each dermatologist's differential. The final differential was thus based on the aggregated "votes" across three board-certified dermatologists.

We verified that this procedure provides substantially higher reproducibility in differential diagnoses than between individual dermatologists (0.73 vs 0.62, see more details in Supplementary Methods). The distribution of the top differential diagnoses is presented in Table 1.

Reference Standard Labeling: Development Set

The development set was further split into a training set to "learn" the neural network weights, and a tuning set to select hyperparameters for the training process. To maximize the amount of training data, more dermatologists labeled the development set: 1-29 dermatologists (from a cohort of 38 U.S. board-certified and 5 Indian board-certified dermatologists) labeled each case. Only cases considered by all of the dermatologists grading that case as having multiple skin conditions or undiagnosable were discarded. Reference standard differential diagnosis was established the same way as for the validation set.

Labeling tool and skin condition mapping

Our labeling tool provided a search-as-you-type interface (see Supplementary Table 5 and Supplementary Fig. 3) based on the standardized Systematized

Nomenclature of Medicine-Clinical Terms (SNOMED-CT)[45], within which more than 20,000 terms were related to cutaneous disease. If the dermatologist could not find a matching SNOMED-CT term, the diagnosis could be entered as free text.

Because SNOMED-CT contains terms at varying granularities and have complex and incompletely-specified relationships between terms[46], three board-certified dermatologists mapped these terms and free text diagnoses entries to a list. The list was initially populated with dermatologic conditions that were common or high-acuity, and more conditions were added as needed. Considerations during this mapping were a granularity that would (1) allow a non-dermatology clinician to reasonably determine the next steps in clinical care, (2) enable clear and concise communication with another healthcare provider, and (3) exclude superfluous information for most purposes (e.g. specific site of the condition). For example, a diagnosis such as "alopecia" would be too broad, but "alopecia areata" and "androgenetic alopecia" would allow a non-dermatologist to engage in next steps in clinical care.

As labels for cases were collected, additional conditions were added to the list as appropriate based on the discussion of at least two of the three dermatologists. Some diagnoses were marked as invalid if they were too broad, non-skin entries, reflected multiple skin conditions (such as a syndrome with multiple skin findings), or were semantically unclear (e.g. tooth abrasion). All mappings were performed while blinded to DLS predictions and the identity of the clinicians or the cases for which the diagnoses were provided. The final list contained 421 conditions (Supplementary Table 6).

Selection of the 26 skin conditions

As in actual clinical practice, the prevalence of difference skin conditions was heavily skewed in our dataset, ranging from skin conditions with >10% prevalence like acne, eczema, and psoriasis, to those with sub-1% prevalence like lentigo, melanoma, and stasis dermatitis. To ensure that there was sufficient data to develop and evaluate the DLS, we filtered the 421 conditions to the top 26 with the highest prevalence based on the development set (when the labeling was approximately 80% complete). Specifically, this ensured that for each of these conditions, there were at least 100 cases in the development dataset (for DLS training purposes), and an projected 25 cases in the validation set (for DLS evaluation). The remaining conditions were aggregated into an "Other" category (which comprised 22% of the cases in validation dataset A).

DLS development

The DLS has two main components, an image-processing deep convolutional neural network, and a shallow network that processes clinical metadata (demographic information and medical history). The image processing component consisted of a variable number (1-6, depending on the number of images in each case) of Inception-v4[47] modules with shared weights. All images were resized to 459×459 pixels, the default size of this network architecture. The clinical metadata were featurized using the one-hot encoding for all categorical features. Age was used as a number normalized to [0,1] based on the range in the development set. These two components were joined at the top using a fully-connected layer (i.e. late fusion[48]).

To help the DLS learn to predict a differential diagnosis (as opposed to a pure classification to predict a single label), the target label of the DLS was based on each case's reference standard differential diagnosis. Specifically, the summed "votes" of each condition in the differential was normalized (to sum to 1), and the DLS was trained using a softmax cross-entropy loss to learn these "soft" target labels. To account for class imbalance, when calculating cross entropy loss, each class was weighted as a function of its frequency, so that cases of rare conditions would contribute more to the loss function. The network weights were optimized using a distributed stochastic gradient descent implementation[49], to predict both the full list of 421 conditions and the shorter list of 27 conditions (26 conditions plus "Other"). To speed up the training and improve training performance, batch normalization[50] and pre-initialization from ImageNet dataset were used[51]. Training was stopped after a fixed number of steps (100,000) with a batch size of 8.

To train the DLS, the development set was partitioned into a training set to learn DLS's parameters, and a tuning set to tune hyperparameters. Because of the severe class imbalance, we created the tune set via stratified sampling (of up to 50 cases per condition). To ensure a clean split with respect to patients, all cases from the patients represented in this sampling were moved to the tune set.

Data augmentation was applied to improve generalization: random flipping, rotating, cropping, and color perturbation. The random cropping was parameterized to ensure that the crops had a minimum overlap of 20% with the pathologic skin region (a separately-collected label for every case in the training set). Random dropout was applied to metadata features (assigned to unknown), to help improve robustness to missing values or potential data errors. Six networks were trained with the same input and hyperparameters (see Supplementary Table 7 for a complete list of hyperparameters), and ensembled[52] to provide the final prediction.

DLS evaluation

To evaluate the DLS performance, we compared its predicted differential diagnosis with the "voted" reference standard differential diagnosis using the top-k accuracy and the average top-k sensitivity. The top-k accuracy measures how frequently the top k predictions capture any of the primary diagnosis in the reference standard (i.e. ranked first in the differential). The top-k sensitivity assesses this for each of the 26 conditions separately, whereas the final average top-k sensitivity is the average across the 26 conditions. Averaging across the 26 conditions avoids biasing towards more common conditions, particularly in validation set A. We use both the top-1 and top-3 metrics in this study.

In addition to comparing both the DLS and clinicians against the voting-based reference standard differential diagnoses, we also evaluated against a reference standard based on agreeing with "at least one" of the three board-certified dermatologists comprising the reference standard ("$Accuracy_{any}$", see Supplementary Tables 2, 8-9).

Finally, we also measured the agreements in the full differential diagnosis between the DLS and reference standard using the average overlap (AO)[37,38]. Because the clinicians were instructed to provide up to three diagnoses, we similarly filtered the DLS's predictions to retain the top-3. Next, unlikely diagnoses lower than a predicted likelihood of 0.1 (selected based on the AO computed on the tune dataset) were filtered to produce the final DLS-predicted differential: up to three diagnoses in ranked order.

Comparison to clinicians

To compare the DLS performance with clinicians, a group of 18 clinicians (who did not participate in prior parts of this study) provided differential diagnoses for validation set B. These clinicians were comprised of three groups of six U.S. board-certified clinicians: dermatologists, PCPs, and NPs. The NPs were selected from those who were practicing independently as primary care providers without physician supervision. Every clinician graded a random one-third of the cases, and each case was graded by two random clinicians from each group (six clinicians total). These clinicians used the same labeling tool as the dermatologists involved in determining the reference standard, and their diagnoses were mapped and processed similarly. In case of ties, the top k diagnoses were determined by randomly selecting the diagnosis from the tied candidates. This tie-breaking affected the top-1 analyses for 13% of dermatologist-provided, 24% of PCP-provided, and 14% of NP-provided diagnoses. The top-3 analysis was minimally affected, with no ties from dermatologists and NPs, and 0.6% ties from PCPs. This tie-breaking avoided confounding the analysis by biasing towards clinicians who provided tied differential diagnoses (which indicates uncertainty).

Feature importance

Additionally, we investigated the relative importance of different types of inputs on the DLS performance. To study the effect of the number of images, we selected a random subset of the images for each case and measured the DLS's performance on this subsampled dataset. For the clinical metadata, we used a permutation procedure ("permutation feature importance"[53]). Briefly, for a metadata variable of interest, this procedure randomly permutes its assignment across cases in the validation set A. Next, the performance of the DLS was measured using the perturbed dataset. To understand the importance of all the metadata collectively, we "dropped out" all the metadata by assigning all their values to unknown. Because the network could have been "dependent" on metadata in this analysis, thus over-representing the importance of metadata, we further trained a DLS using only images, and evaluated its performance. Finally, we used integrated gradients[39] to highlight the parts of each image that have the biggest effect on the prediction.

Statistical analysis

To compute the confidence intervals (CIs), we used a non-parametric bootstrap procedure[54] with 1,000 samples. Because of the intensive compute required to re-run DLS inference, CIs for the feature importance analyses were calculated using the normal approximation with 20 runs (1.96×standard error, with each run performed on the entire validation set A). To compare the DLS performance to clinicians, a standard permutation test[55] was used. Briefly, in each of the 10,000 trials, the DLS's score was randomly swapped with itself or a comparator clinician's score for each case, yielding a DLS-human difference in top-1 accuracy sampled from the null distribution. To perform the non-inferiority test, the empirical p-value was computed by adding the 5% margin to the observed difference and comparing this number to its empirical quantiles[54,55]. Non-inferiority compared to dermatologists in top-1 accuracy was documented in an institutional mailing list as our pre-specified primary endpoint prior to evaluating the DLS on the validation dataset.

# FIGURES

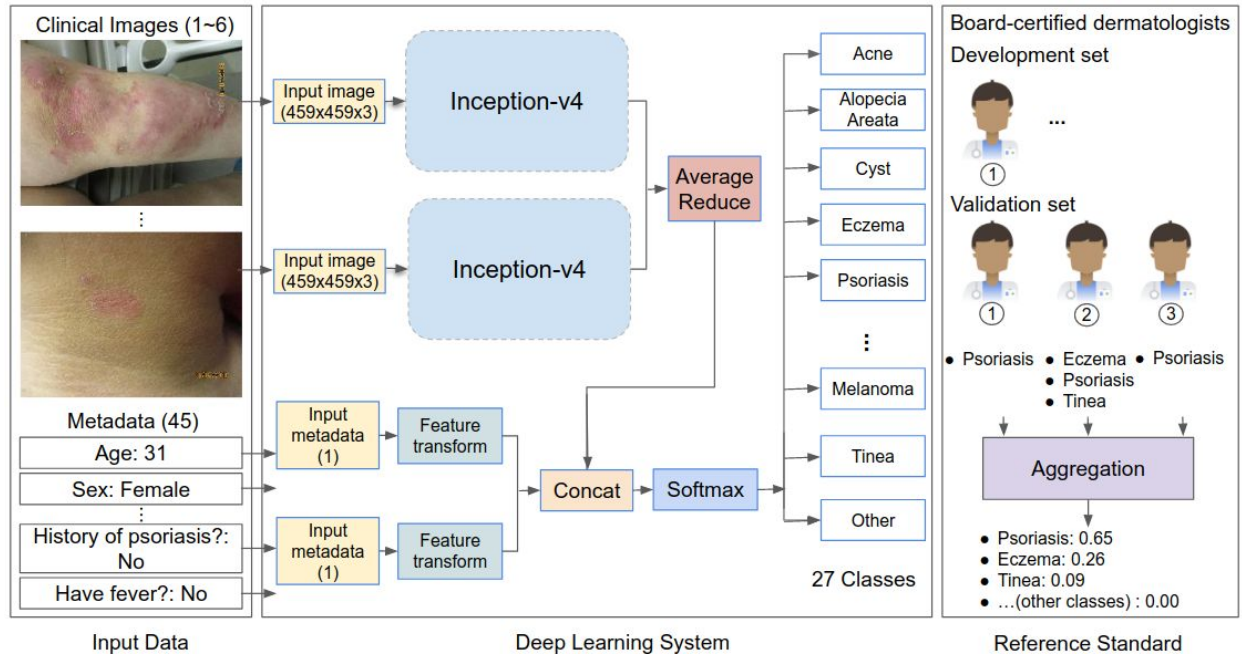

**Fig. 1 | Overview of the development and validation of our deep learning system (DLS).**
For each case, the DLS takes as input 1 to 6 de-identified skin photographs and 45 metadata variables such as demographic information and medical history (left). The DLS then processes the images using Inception-v4 modules with shared weights before applying an average pool and concatenating with the metadata features. The output of the classification layer of the DLS is the relative likelihood of 27 categories (26 skin conditions plus "Other", Table 1). These conditions were chosen based on a granularity that could guide a non-dermatologist clinician to next steps in clinical care. The labels used to develop and validate the DLS were provided by board-certified dermatologists; one or more dermatologists per case for training, and three dermatologists per case for the validation set. For each case, each dermatologist provided their top three differential diagnosis. The multiple differential diagnoses are then aggregated into a single ranked list (see Supplementary Fig. 8). During training, the aggregated ranked list of dermatologist-provided diagnoses have an associated aggregated "confidence" score per diagnosis, and these confidences are the target "soft" labels for the DLS. The DLS therefore learns from both the primary (top-ranked) diagnosis as well as the lower-ranked diagnoses. In this way, the DLS was trained to provide a differential diagnosis instead of a single prediction output.

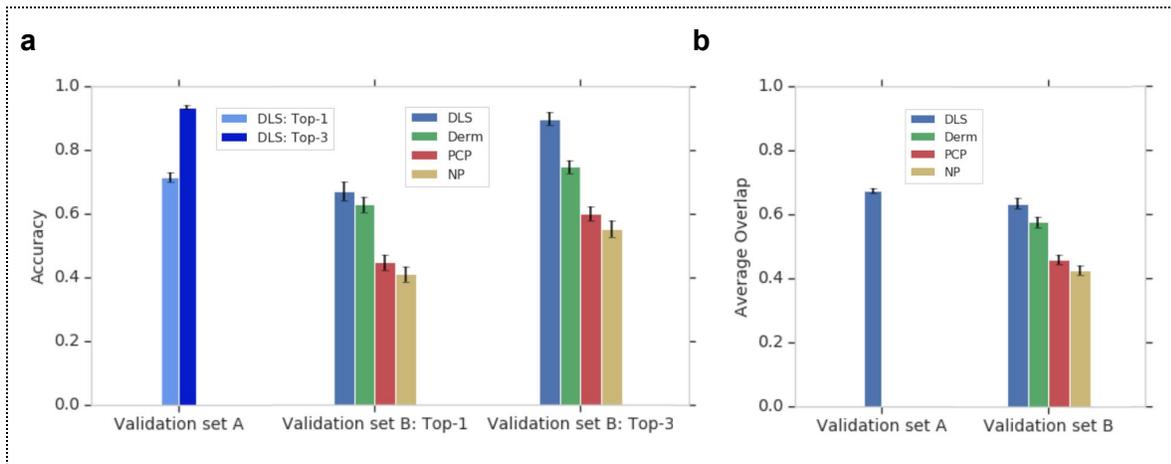

**Fig. 2 | Performance of the deep learning system (DLS) and the clinicians: dermatologists (Derm), primary care physicians (PCP), and nurse practitioners (NP) on validation set A and B. a**, Top-1 and top-3 accuracy for the DLS and clinicians. The sensitivity of the DLS for each of the 26 conditions is presented in Extended Data Fig. 1. **b**, Average overlap (to assess the full differential diagnosis) of the DLS and clinicians. Average overlap ranges from 0 to 1, with higher values indicating better agreement. Error bars indicate 95% confidence intervals.

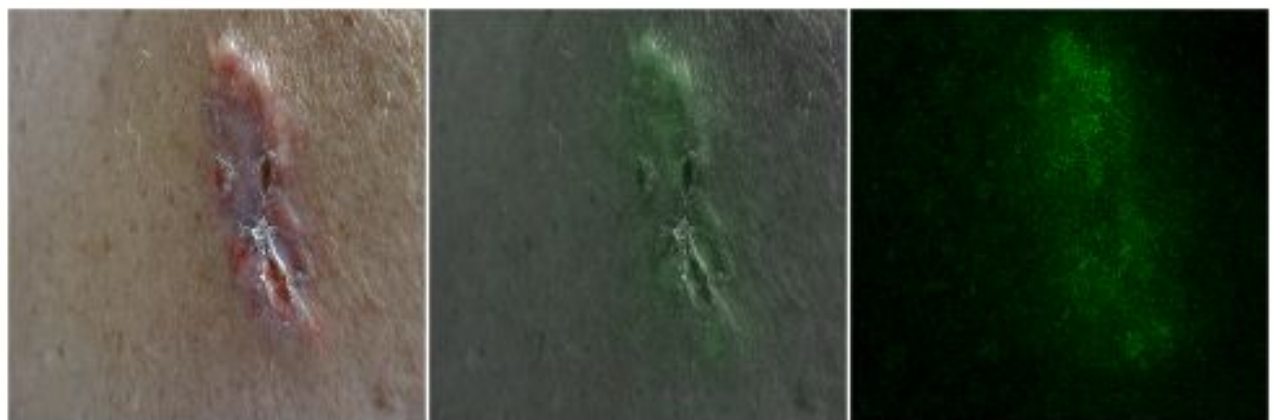

| Reference standard | DLS (top 3) | DLS (growth subgroup) | NP (missed) | NP (missed) | PCP (tied 1st diagnosis) | PCP (missed) | Derm | Derm |
|---|---|---|---|---|---|---|---|---|
| BCC; SCC/SCCIS; Scar condition | BCC: 0.84; Scar condition: 0.06; SCC/SCCIS: 0.05 | Malignant: 1.0; Benign: 0.0 | Other (hypertrophic skin); Scar condition | AK; Other (skin lesion); Psoriasis | BCC / SCC/SCCIS; Melanoma | Psoriasis | BCC | BCC |

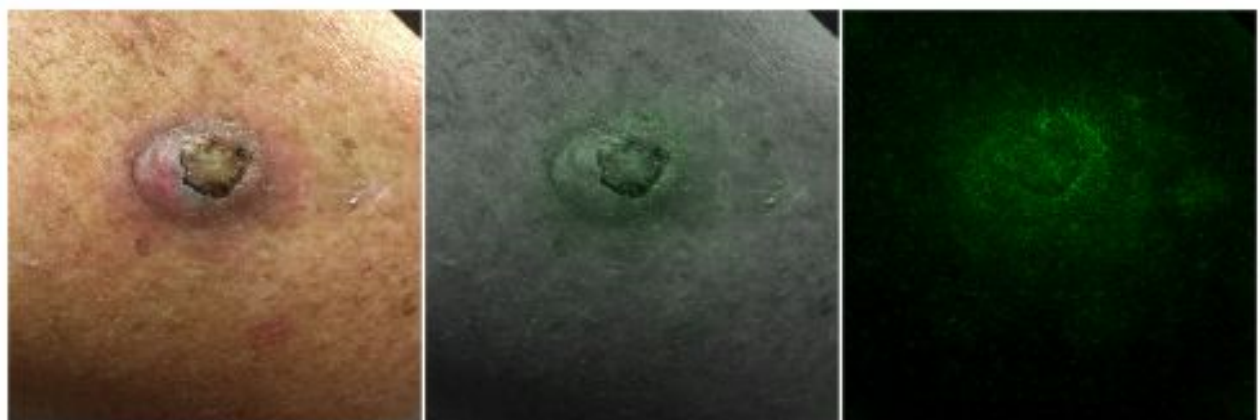

| Reference standard | DLS (top 3) | DLS (growth subgroup) | NP (2nd diagnosis) | NP (tied 1st diagnosis) | PCP (missed) | PCP (missed) | Derm | Derm |
|---|---|---|---|---|---|---|---|---|
| SCC/SCCIS; BCC | SCC/SCCIS: 0.74; BCC: 0.19; Actinic keratosis: 0.04 | Malignant: 0.94; Benign: 0.06 | BCC; SCC/SCCIS; Melanoma | Other (skin lesion) / SCC/SCCIS; BCC | Cannot diagnose | Other (pyoderma) | SCC/SCCIS; BCC | SCC/SCCIS |

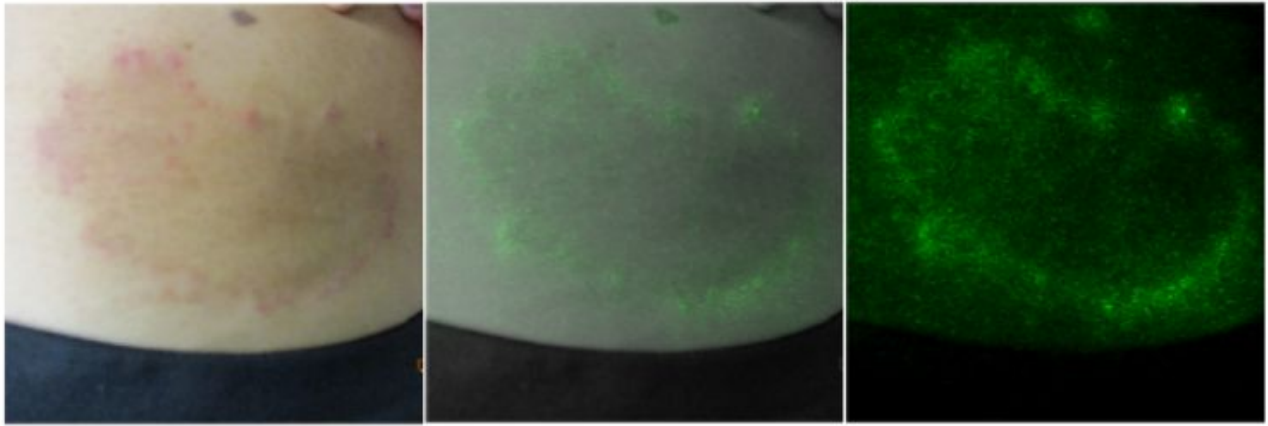

c

| Reference standard | DLS (top 3) | DLS (erythematosquamous and papulosquamous subgroup) | NP (missed) | NP (missed) | PCP (tied 1st diagnosis) | PCP (missed) | Derm | Derm |
|---|---|---|---|---|---|---|---|---|
| Tinea | Tinea: 0.95; Other: 0.03; Eczema: 0.02 | Infectious: 0.98; Non-infectious: 0.02 | Eczema / Other (Chronic contact dermatitis); Psoriasis | Other (Generalized granuloma annulare) | Other (Granuloma annulare) / Tinea | Eczema | Tinea; Other (Granuloma annulare) | Tinea |

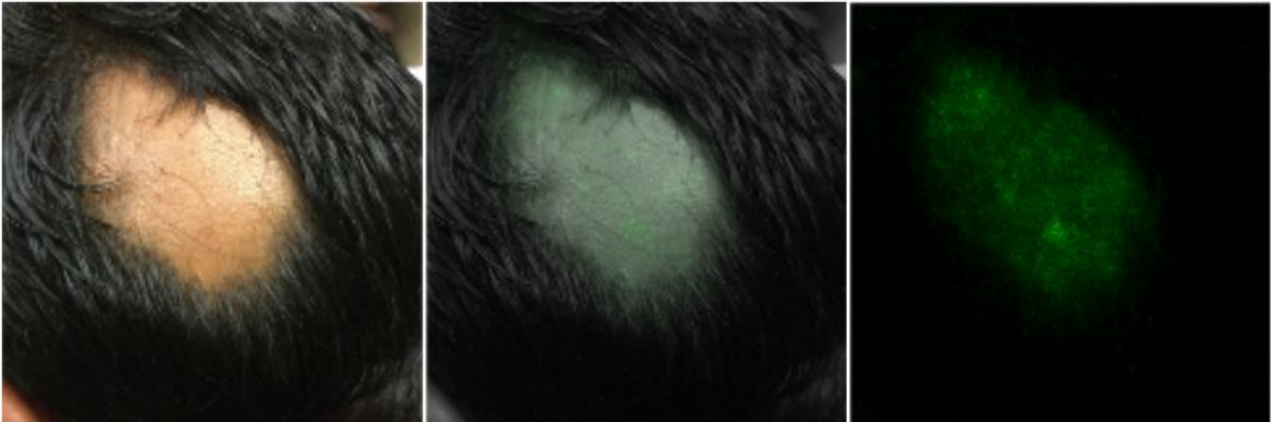

d

| Reference standard | DLS (top 3) | DLS (hair loss subgroup) | NP (3rd diagnosis) | NP (missed) | PCP (2nd diagnosis) | PCP | Derm | Derm |
|---|---|---|---|---|---|---|---|---|
| AA | AA: 0.89; Other: 0.05; AGA: 0.03 | AA: 0.97; AGA: 0.03 | AGA; Other (Alopecia localis); AA | AGA | AGA; AA | AA | AA | AA; Other (trichotillomania) |

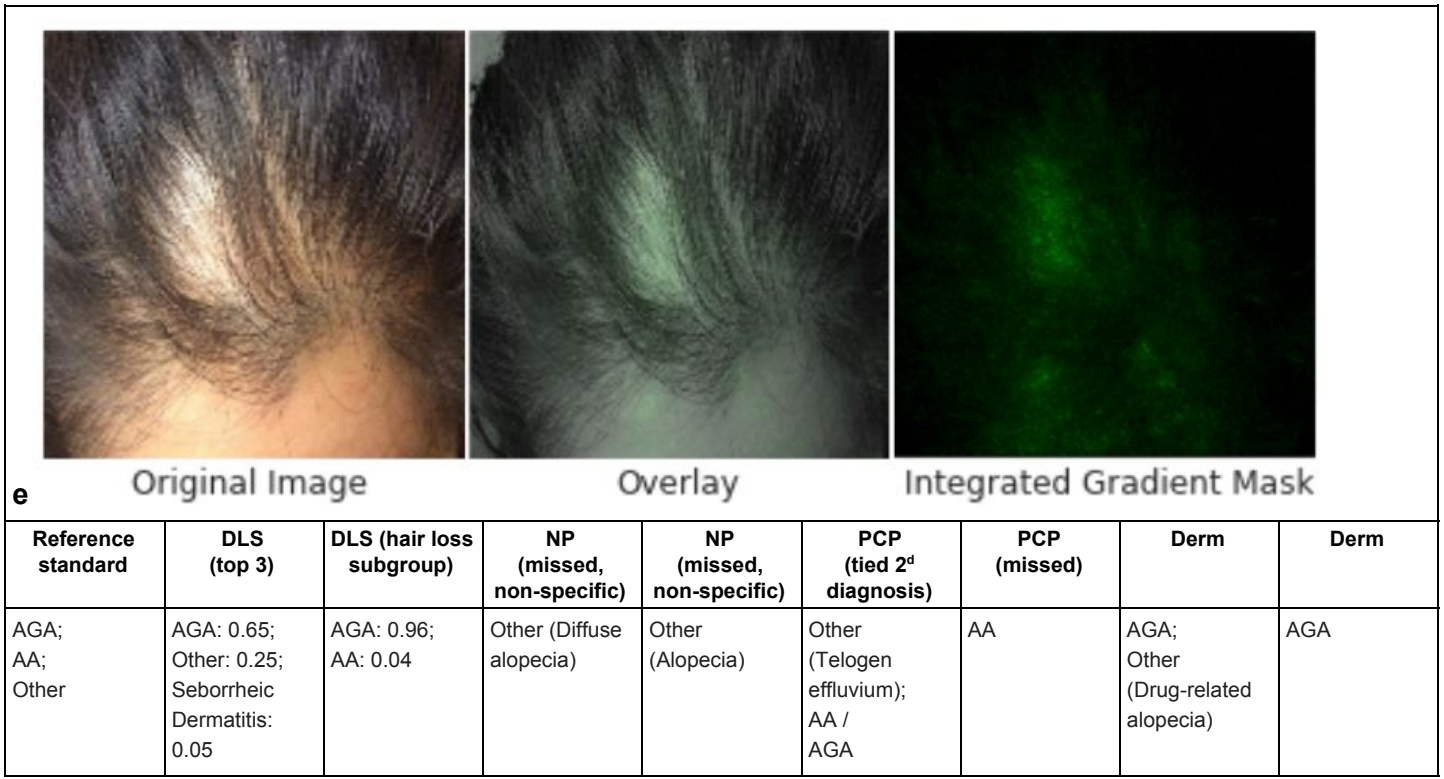

| Reference standard | DLS (top 3) | DLS (hair loss subgroup) | NP (missed, non-specific) | NP (missed, non-specific) | PCP (tied 2ᵈ diagnosis) | PCP (missed) | Derm | Derm |
|---|---|---|---|---|---|---|---|---|
| AGA; AA; Other | AGA: 0.65; Other: 0.25; Seborrheic Dermatitis: 0.05 | AGA: 0.96; AA: 0.04 | Other (Diffuse alopecia) | Other (Alopecia) | Other (Telogen effluvium); AA / AGA | AA | AGA; Other (Drug-related alopecia) | AGA |

**Fig. 3 | Representative examples of challenging cases missed by non-dermatologists.** For each case, an original image is provided on the left, and with a saliency mask on the right. The middle image shows the original image in grayscale, with the saliency overlaid in green. All clinicians were instructed to be as specific as possible when providing the diagnostic labels. Diagnoses for the reference standard and comparator clinicians who reviewed each case are included here and ranked by confidence from top to bottom. **a**, The DLS's primary diagnosis of basal cell carcinoma (BCC) concurs with the reference standard, both comparator dermatologists, and one PCP. Both NPs and one PCP missed this diagnosis. **b**, The DLS's primary diagnosis of squamous cell carcinoma (SCC/SCCIS) concurs with the reference standard and both comparator dermatologists. Both NPs considered another diagnosis as more or equally likely, while the PCPs missed this diagnosis. **c**, The DLS's primary diagnosis of tinea concurs with the reference standard and primary diagnoses of the comparator dermatologists. One PCP considered another diagnosis as equally likely, while the other PCP and both NPs missed this diagnosis. **d**, The DLS, comparator dermatologists, and one PCP all agreed with the reference standard of Alopecia Areata (AA). This was missed as a primary diagnosis by both NPs and one of the PCPs. **e**, The DLS's primary diagnosis of Androgenetic Alopecia (AGA) concurs with the reference standard and both comparator dermatologists. This was missed as a primary diagnosis by both NPs and both PCPs. In the last two cases (panels d and e), diagnosing the specific type of alopecia is important because AGA and AA have different treatments. More details about these cases are presented in Supplementary Fig. 9.

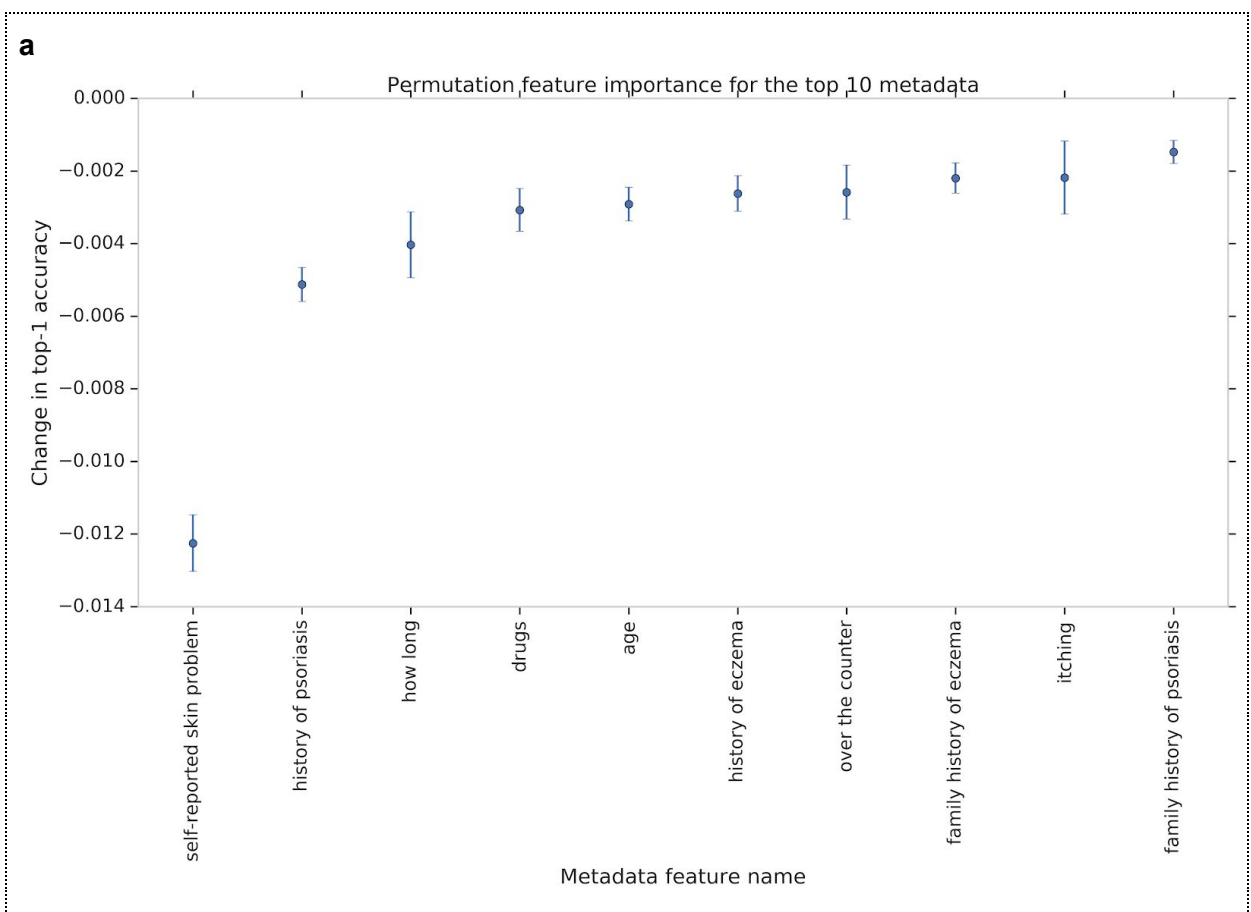
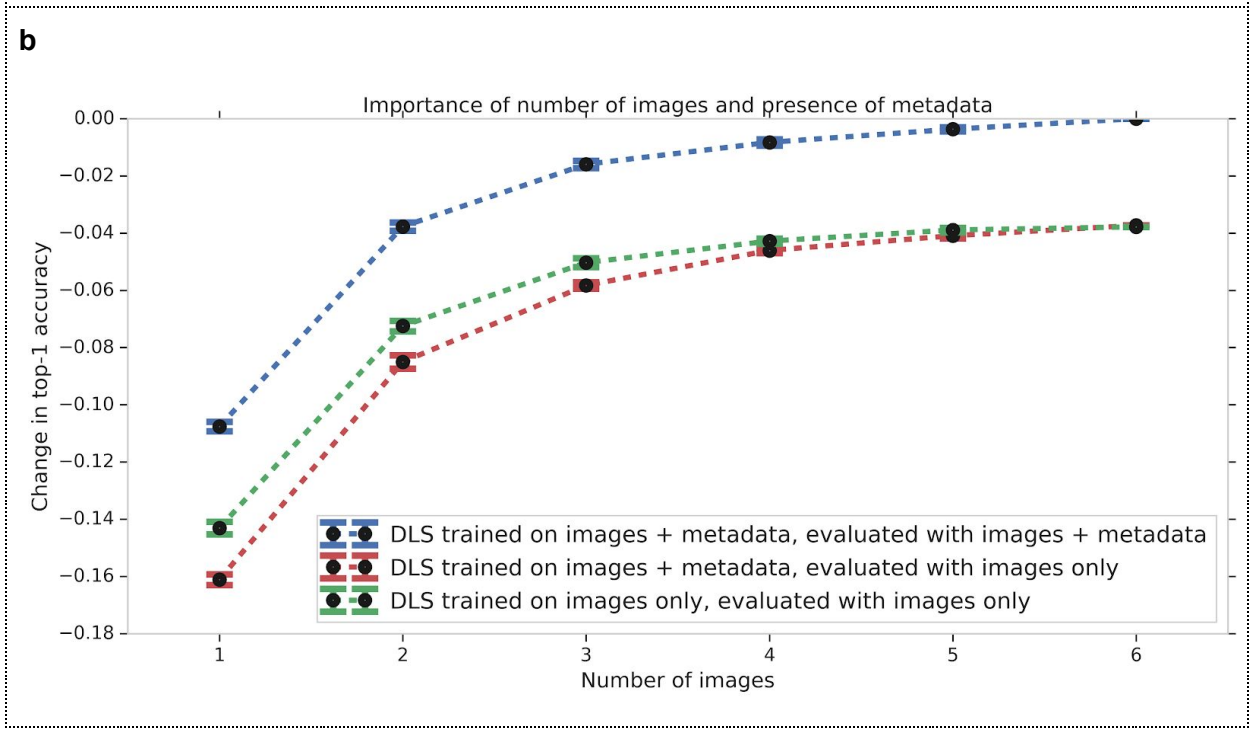

**Fig. 4 | Importance of different inputs to the deep learning system (DLS). a**, Impact on the top-1 accuracy of permuting each of the top 10 most important clinical metadata across validation set A examples, using the same trained DLS (for all metadata, see Supplementary Fig. 6). **b**, The blue line illustrates the impact on the top-1 accuracy of different numbers of input images for the same DLS (that was trained using all images and metadata). The red line illustrates a similar trend when the clinical metadata are absent from this same DLS. Finally, the green line illustrates the trend, but for a DLS retrained without using clinical metadata (so that the DLS cannot depend on the presence of clinical metadata). All trends were the average of 20 different runs to reduce the effects of stochasticity from the permutation, image sampling, and/or training process. The error bars indicate 95% confidence intervals.

# TABLES

**Table 1 | Dataset characteristics.** The dataset contained clinical cases from a teledermatology practice serving 17 primary care and specialist sites from 2 states in the U.S.. To mimic a prospective design, the dataset was split temporally into a development set (cases seen between 2010 and 2017) and validation set A (cases seen between 2017 and 2018). Validation set B was a subset of set A that was enriched for rarer skin conditions in this study, and was reviewed by three groups of clinicians for comparison.

| **Characteristics** | Development set | Validation set A | Validation set B (subset of "A") |
|---|---|---|---|
| **Years** | 2010 to 2017 | 2017 to 2018 | 2017 to 2018 |
| **Total no. of cases** | 16,539 | 4,145 | N/A |
| **No. of cases with multiple skin conditions (excluded from study)** | 1,394 | 224 | N/A |
| **No. of cases indicated as not-diagnosable by dermatologists (excluded from study)** | 1,124 | 165 | N/A |
| **No. of cases included in study** | 14,021 | 3,756 | 963 |
| **No. of images included in study** | 56,134 | 14,883 | 3,707 |
| **No. of patients included in study** | 11,026 | 3,241 | 933 |
| **Age\*, median (25th, 75th percentiles)** | 40 (27, 54) | 40 (28, 54) | 43 (30, 56) |
| **Female (%)** | 8,637 (61.6%) | 2,371 (63.1%) | 615 (63.9%) |
| **Fitzpatrick skin types (6 types)\*\*** <br>  Type I (%) <br>  Type II (%) <br>  Type III (%) <br>  Type IV (%) <br>  Type V (%) <br>  Type VI (%) <br>  Unknown (%) | <br>36 (0.3%)<br>2,419 (17.3%)<br>5,768 (41.1%)<br>4,457 (31.8%)<br>456 (3.3%)<br>41 (0.3%)<br>844 (6.0%) | <br>9 (0.2%)<br>383 (10.2%)<br>2412 (64.2%)<br>724 (19.3%)<br>101 (2.7%)<br>1 (0.0%)<br>126 (3.4%) | <br>0 (0.0%)<br>104 (10.8%)<br>607 (63.0%)<br>195 (20.2%)<br>24 (2.5%)<br>0 (0.0%)<br>33 (3.4%) |
| **Skin conditions based on primary diagnosis (26 conditions, plus "other")\*\*\*** <br>  Acne (%) <br>  Actinic Keratosis (%) | <br><br>1,512 (10.8%)<br>167 (1.2%) | <br><br>407 (10.8%)<br>49 (1.3%) | <br><br>40 (4.2%)<br>34 (3.6%) |
| Allergic Contact Dermatitis (%) | 153 (1.1%) | 36 (0.9%) | 25 (2.6%) |
| Alopecia Areata (%) | 290 (2.1%) | 96 (2.5%) | 37 (3.8%) |
| Androgenetic Alopecia (%) | 135 (1.0%) | 50 (1.3%) | 33 (3.4%) |
| Basal Cell Carcinoma (%) | 242 (1.7%) | 45 (1.2%) | 28 (2.9%) |
| Cyst (%) | 236 (1.7%) | 86 (2.3%) | 31 (3.2%) |
| Eczema (%) | 1,987 (14.2%) | 659 (17.5%) | 50 (5.2%) |
| Folliculitis (%) | 273 (1.9%) | 87 (2.3%) | 32 (3.3%) |
| Hidradenitis (%) | 149 (1.1%) | 45 (1.2%) | 35 (3.6%) |
| Lentigo (%) | 86 (0.6%) | 33 (0.9%) | 32 (3.3%) |
| Melanocytic Nevus (%) | 656 (4.7%) | 183 (4.9%) | 35 (3.6%) |
| Melanoma (%) | 84 (0.6%) | 22 (0.6%) | 19 (1.9%) |
| Post Inflammatory Hyperpigmentation (%) | 142 (1.0%) | 51 (1.4%) | 29 (3.0%) |
| Psoriasis (%) | 1,843 (13.1%) | 335 (8.9%) | 39 (4.1%) |
| Squamous Cell Carcinoma / Squamous Cell Carcinoma In Situ (SCC/SCCIS) (%) | 128 (0.9%) | 36 (1.0%) | 33 (3.5%) |
| Seborrheic Keratosis / Irritated Seborrheic Keratosis (SK/ISK) (%) | 612 (4.4%) | 211 (5.6%) | 38 (4.0%) |
| Scar Condition (%) | 275 (2.0%) | 60 (1.6%) | 33 (3.4%) |
| Seborrheic Dermatitis (%) | 286 (2.0%) | 98 (2.6%) | 37 (3.8%) |
| Skin Tag (%) | 213 (1.5%) | 70 (1.9%) | 33 (3.4%) |
| Stasis Dermatitis (%) | 103 (0.7%) | 26 (0.7%) | 25 (2.6%) |
| Tinea (%) | 213 (1.5%) | 34 (0.9%) | 31 (3.2%) |
| Tinea Versicolor (%) | 182 (1.3%) | 36 (0.9%) | 35 (3.6%) |
| Urticaria (%) | 116 (0.8%) | 34 (0.9%) | 33 (3.4%) |
| Verruca Vulgaris (%) | 343 (2.4%) | 83 (2.2%) | 34 (3.5%) |
| Vitiligo (%) | 200 (1.4%) | 74 (2.0%) | 36 (3.7%) |
| Other (%) | 3,395 (24.2%) | 813 (21.6%) | 98 (10.2%) |

\* Ages were truncated at 90 as part of the de-identification process. For each dataset, the minimum age was 18 and the maximum age was 90.

\*\* Fitzpatrick skin type was obtained via the majority opinion of three raters trained by dermatologists to distinguish skin types. Some cases' skin types were labeled as "unknown" because of reasons such as lack of majority agreement among raters, inconsistent skin types observed in different images, and insufficient visible skin regions.

\*\*\* When multiple primary diagnosis exist, the contribution of each condition in the list towards its total count was fractionalized, such that the total number of cases over all conditions sums up to the size of each dataset. This causes a slight difference when compared to the numbers as part of the x-axes labels in Extended Data Fig. 1, where each condition was treated independently.

**Table 2 | Sensitivity of the deep learning system (DLS) and three types of clinicians (dermatologists, Derm; primary care physicians, PCP; and nurse practitioners, NP) for clinically relevant and challenging subgroups based on appearance on clinical presentation.** Bold indicates the highest value for each subcategory and each evaluation metric.

| Clinically relevant groups | Subcategory | Conditions in the subcategory | No. of cases | Top-1 Sensitivity | | | | Top-3 Sensitivity | | | |
|---|---|---|---|---|---|---|---|---|---|---|---|
| | | | | DLS | Derm | PCP | NP | DLS | Derm | PCP | NP |
| Growth | Malignant | Basal cell carcinoma, melanoma, squamous cell carcinoma / squamous cell carcinoma in situ (SCC/SCCIS) | 83 | 0.57 | **0.71** | 0.55 | 0.53 | 0.88 | **0.89** | 0.69 | 0.72 |
| | Benign | Actinic keratosis, cyst, lentigo, melanocytic nevus, seborrheic keratosis / irritated seborrheic keratosis (SK/ISK), skin tag, verruca vulgaris | 184 | **0.75** | 0.64 | 0.54 | 0.42 | **0.92** | 0.76 | 0.69 | 0.59 |
| Erythematosquamous and papulosquamous skin disease | Infectious | Tinea, Tinea versicolor | 71 | **0.75** | 0.68 | 0.51 | 0.48 | **0.91** | 0.85 | 0.70 | 0.60 |
| | Non-infectious | Eczema, Psoriasis, Stasis dermatitis, Allergic contact dermatitis, Seborrheic dermatitis | 202 | **0.67** | 0.43 | 0.49 | 0.46 | **0.95** | 0.55 | 0.62 | 0.57 |
| Hair loss | Alopecia areata | Alopecia areata | 39 | 0.77 | **0.80** | 0.59 | 0.45 | 0.86 | **0.91** | 0.77 | 0.64 |
| | Androgenetic alopecia | Androgenetic alopecia | 34 | **0.79** | 0.69 | 0.28 | 0.22 | **0.91** | 0.84 | 0.43 | 0.37 |

Author contributions and acknowledgements

Author contribution

Y.L.(1), A.J., C.E., D.H.W., K.L., and D.C prepared the dataset for usage. S.H., K.K., and R.H.-W. provided clinical expertise and guidance for the study. Y.L.(1), A.J., C.E., K.L., P.B., G.O.M., J.G., D.A., S.H., and K.K. worked on the technical, logistical, and quality control aspects of label collection. S.H. and K.K. established the skin condition mapping. Y.L.(1), K.L., V.G., and D.C. developed the model; Y.L.(1), A.J., N.S., and V.N. performed statistical analysis and additional analysis. Y.L.(2) guided study design, analysis of the results, and statistical analysis. S.G. studied the potential utility of the model. R.C.D. and D.C. initiated the project and led the overall development, with strategic guidance and executive support from G.S.C., L.H.P., and D.R.W.. Y.L.(1), Y.L.(2), and S.H. prepared the manuscript with the assistance and feedback from all other co-authors.

Competing interests

K.K. and S.H. were consultants of Google LLC. R.H.-W. is an employee of the Medical University of Graz. G.O.M. is an employee of Adecco Staffing supporting Google LLC. This study was funded by Google LLC. The remaining authors are employees of Google LLC and own Alphabet stock as part of the standard compensation package. The authors have no other competing interests to disclose.

Acknowledgement

The authors would like to acknowledge William Chen, Jessica Yoshimi, Xiang Ji, and Quang Duong for software infrastructure support for data collection. Thanks also go to Genevieve Foti, Ken Su, T Saensuksopa, Devon Wang, Yi Gao, and Linh Tran. Last but not least, this work would not have been possible without the participation of the dermatologists, primary care physicians, and nurse practitioners who reviewed cases for this study, and Sabina Bis who helped to establish the skin condition mapping, and Amy Paller who provided feedback on the manuscript.

# Extended Data

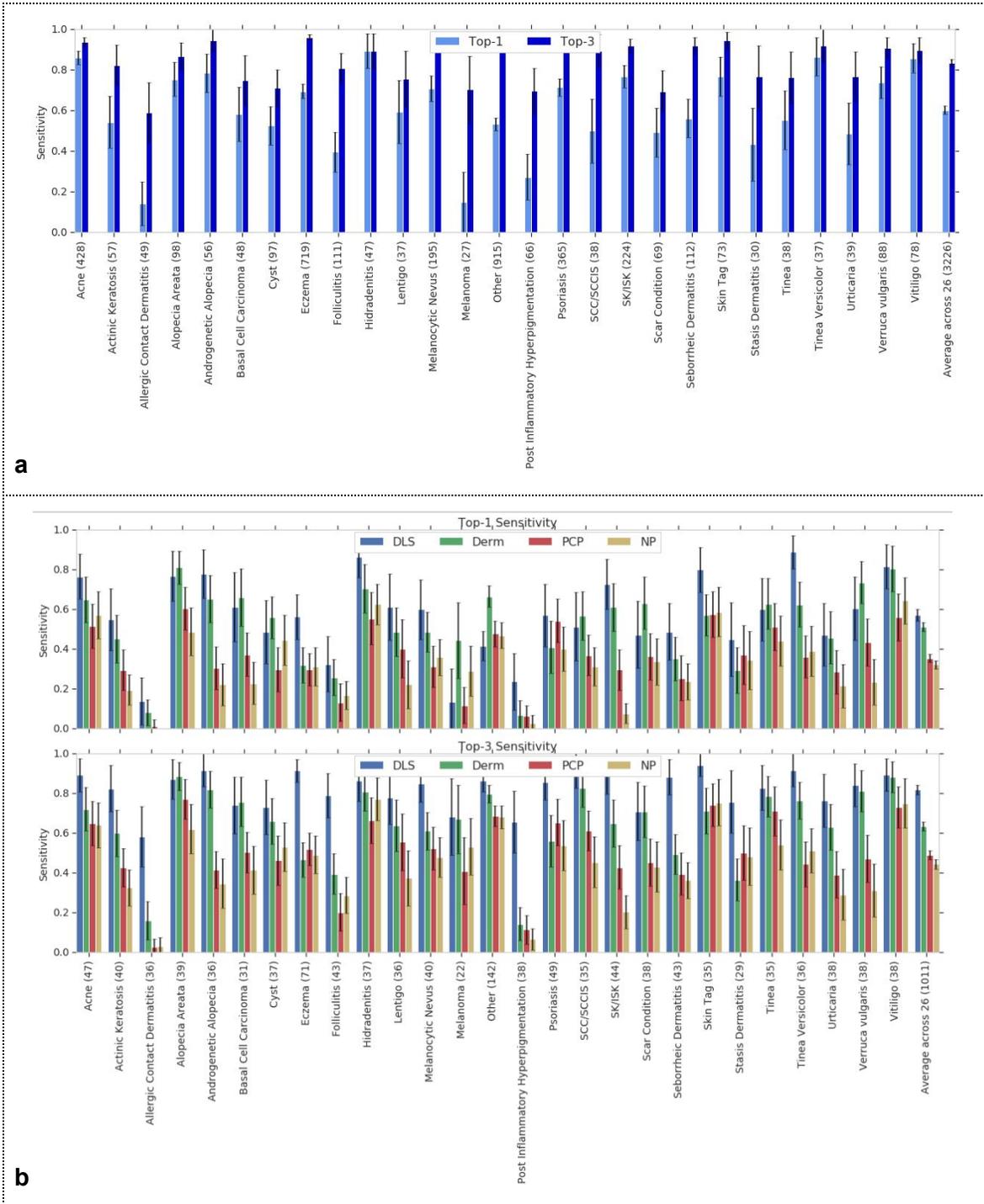

**Extended Data Fig. 1 | Performance of the deep learning system (DLS) and clinicians, broken down for each of the 26 categories of skin conditions. a**, Top-1 and top-3 sensitivity of the DLS on validation set A. **b**, Top-1 and top-3 sensitivity of the DLS and three types of

clinicians: dermatologists (Derm), primary care physicians (PCP), and nurse practitioners (NP). Numbers in parentheses in the x-axes indicate the number of cases. Detailed breakdown of each clinician and the DLS performance on the subset of cases graded by each clinician are in Supplementary Table 2. Error bars indicate 95% CI.

**Extended Data Table 1 | Performance of the deep learning system (DLS) and different types of clinicians, on validation sets A and B.** The reference standard differential diagnoses for each case was determined by the votes of a panel of three board-certified dermatologists. Performance was measured by the agreement of the top-1 and top-3 diagnoses with the primary diagnosis of the panel. The average overlap (AO) directly compares the DLS or clinician-provided ranked differential diagnoses with the panel's full differential diagnoses. The AO ranges from 0 to 1, with higher values indicating better agreement. Numbers in square braces indicate 95% confidence intervals. Bold indicates the highest value within each column for validation set B.

| Dataset | Grader | Top-1 | | Top-3 | | Average Overlap (AO) |
|---|---|---|---|---|---|---|
| | | Accuracy | Average Sensitivity | Accuracy | Average Sensitivity | |
| Validation set A | DLS | 0.71 [0.70, 0.73] | 0.60 [0.58, 0.62] | 0.93 [0.93, 0.94] | 0.83 [0.82, 0.85] | 0.67 [0.66, 0.68] |
| Validation set B (enriched subset of set A) | DLS | **0.67 [0.64, 0.70]** | **0.57 [0.54, 0.60]** | **0.90 [0.88, 0.92]** | **0.82 [0.80, 0.84]** | **0.63 [0.62, 0.65]** |
| | Derm | 0.63 [0.60, 0.65] | 0.51 [0.49, 0.54] | 0.75 [0.72, 0.77] | 0.64 [0.61, 0.66] | 0.58 [0.56, 0.59] |
| | PCP | 0.45 [0.42, 0.47] | 0.35 [0.33, 0.38] | 0.60 [0.57, 0.62] | 0.49 [0.47, 0.51] | 0.46 [0.44, 0.47] |
| | NP | 0.41 [0.38, 0.43] | 0.32 [0.30, 0.34] | 0.55 [0.52, 0.58] | 0.44 [0.42, 0.47] | 0.43 [0.41, 0.44] |

**Extended Data Table 2 | Performance of the deep learning system (DLS), sliced by self-reported demographic information (including age, sex, race and ethnicity), and Fitzpatrick skin type on validation set A.** Metrics used are identical to the ones in Extended Data Table 1. Numbers in square braces indicate 95% confidence intervals.

| Break down | Category | Top-1 | | Top-3 | | Average Overlap (AO) |
|---|---|---|---|---|---|---|
| | | Accuracy | Average Sensitivity | Accuracy | Average Sensitivity | |
| Age | [18, 30) (29.5%) | 0.77 [0.75, 0.80] | 0.59 [0.55, 0.65] | 0.95 [0.93, 0.96] | 0.80 [0.77, 0.87] | 0.71 [0.70, 0.72] |
| | [30, 40) (19.9%) | 0.70 [0.67, 0.74] | 0.53 [0.49, 0.61] | 0.93 [0.91, 0.95] | 0.80 [0.75, 0.85] | 0.66 [0.64, 0.68] |
| | [40, 50) (17.3%) | 0.69 [0.65, 0.72] | 0.59 [0.53, 0.64] | 0.93 [0.91, 0.95] | 0.83 [0.78, 0.87] | 0.66 [0.64, 0.68] |
| | [50, 60) (18.6%) | 0.69 [0.66, 0.73] | 0.63 [0.57, 0.67] | 0.93 [0.91, 0.95] | 0.81 [0.76, 0.85] | 0.66 [0.64, 0.67] |
| | [60, 90] (14.6%) | 0.67 [0.63, 0.71] | 0.50 [0.44, 0.55] | 0.93 [0.91, 0.95] | 0.81 [0.76, 0.87] | 0.64 [0.62, 0.66] |
| Sex | Female (63.1%) | 0.72 [0.70, 0.74] | 0.61 [0.58, 0.64] | 0.94 [0.93, 0.95] | 0.84 [0.81, 0.86] | 0.67 [0.66, 0.68] |
| | Male (36.9%) | 0.71 [0.68, 0.73] | 0.59 [0.55, 0.63] | 0.93 [0.91, 0.94] | 0.83 [0.80, 0.86] | 0.67 [0.66, 0.69] |
| Race and ethnicity | American Indian or Alaska Native (1.1%) | 0.62 [0.48, 0.76] | 0.51** [0.38, 0.71] | 0.93 [0.83, 1.00] | 0.88** [0.75, 1.00] | 0.67 [0.60, 0.74] |
| | Asian (12.6%) | 0.74 [0.70, 0.78] | 0.59 [0.53, 0.65] | 0.95 [0.93, 0.97] | 0.85 [0.79, 0.90] | 0.67 [0.65, 0.70] |
| | Black or African American (6.1%) | 0.73 [0.67, 0.79] | 0.56 [0.50, 0.68] | 0.92 [0.89, 0.95] | 0.74 [0.68, 0.86] | 0.68 [0.65, 0.71] |
| | Hispanic or Latino (43.4%) | 0.72 [0.70, 0.74] | 0.58 [0.54, 0.61] | 0.94 [0.93, 0.95] | 0.83 [0.80, 0.87] | 0.68 [0.67, 0.69] |
| | Native Hawaiin or Pacific Islander (1.6%) | 0.66 [0.54, 0.77] | 0.57 [0.41, 0.70] | 0.98 [0.95, 1.00] | 0.80 [0.68, 0.90] | 0.62 [0.56, 0.67] |
| | White (31.3%) | 0.70 [0.68, 0.73] | 0.60 [0.56, 0.64] | 0.92 [0.90, 0.93] | 0.81 [0.77, 0.84] | 0.66 [0.65, 0.68] |
| | Not specified (3.9%) | 0.68 [0.61, 0.76] | 0.63 [0.53, 0.73] | 0.95 [0.91, 0.98] | 0.85 [0.77, 0.93] | 0.65 [0.62, 0.68] |
| Fitzpatrick skin type | Type I (0.2%) | 0.56 [0.22, 0.89] | 0.67** [0.50, 1.00] | 0.67 [0.33, 0.89] | 0.83** [0.67, 1.00] | 0.52 [0.33, 0.71] |
| | Type II (10.2%) | 0.69 [0.64, 0.74] | 0.62 [0.54, 0.70] | 0.91 [0.89, 0.94] | 0.80 [0.73, 0.86] | 0.66 [0.63, 0.68] |
| | Type III (64.2%) | 0.72 [0.70, 0.74] | 0.61 [0.58, 0.63] | 0.94 [0.93, 0.95] | 0.84 [0.82, 0.87] | 0.67 [0.67, 0.68] |
| | Type IV (19.3%) | 0.71 [0.68, 0.75] | 0.56 [0.49, 0.61] | 0.93 [0.91, 0.94] | 0.79 [0.72, 0.84] | 0.68 [0.66, 0.70] |
| | Type V (2.7%) | 0.79 [0.71, 0.87] | 0.68** [0.60, 0.82] | 0.94 [0.89, 0.98] | 0.80** [0.74, 0.95] | 0.68 [0.63, 0.72] |
| | Type VI (0.0%) | 1.00* | 1.00* ** | 1.00* | 1.00* ** | 0.75* |
| | Unknown (3.4%) | 0.67 [0.57, 0.75] | 0.55 [0.45, 0.70] | 0.95 [0.91, 0.99] | 0.82 [0.73, 0.94] | 0.65 [0.60, 0.70] |

* : There was only 1 case labeled as Type VI, so confidence intervals were not meaningful.
**: At least ten of the 26 conditions were absent from this subanalysis, resulting in an ill-defined sensitivity for those conditions and an unreliable estimate for average sensitivity.

# Supplementary Information

## Supplementary Methods

### Labeler onboarding and certification

In addition to formal board certification, all study participants (dermatologists, PCPs, and NPs) underwent an onboarding process to familiarize with the grading tools. In particular, the dermatologists comprising the reference standard graded 147 cases randomly sampled from the development set as an assessment to ensure consistent grading. For each case, their leading diagnosis was compared to the aggregated opinion of a panel of three experienced U.S. board-certified dermatologists, and only dermatologists who had an top-3 accuracy exceeding 60% participated in determining the reference standard for the validation set. This threshold was chosen based on the statistics of dermatologist grader accuracy, so as to leave room for disagreement in complex cases while ensuring a minimum consistency in grading following guidelines (e.g. specificity of diagnoses) and familiarity with the tool.

### Reference standard voting procedure and reproducibility

Here, we detail the voting procedure[44] used to improve reproducibility of our reference standard (Supplementary Fig. 8). First, each dermatologist provided up to three differential diagnoses and accompanying confidences values in the range [1, 5] for each of the diagnosis. Next, each diagnosis was mapped to a condition. If duplicates occurred (i.e. multiple diagnoses were mapped to the same condition), the highest confidence was retained. The relative ranks of the mapped conditions were used to rank the conditions into a differential diagnosis (i.e. primary, secondary, and tertiary diagnosis). Each mapped condition was then assigned a weight: the inverse of the rank. If multiple mapped conditions shared the same confidence, then the weight was evenly distributed across the conditions. Answers from the dermatologists were then aggregated to form the reference standard, by summing up the weights, before limiting the skin condition classes to 27 (26 conditions plus "Other") and normalizing their weights to sum to 1. Distribution of the number of conditions in the differential diagnosis for each set is shown in Supplementary Fig. 4. Detailed analysis of the secondary and tertiary diagnoses that are provided alongside every primary diagnosis is shown in Supplementary Fig. 5.

To investigate the reproducibility of the reference standard, for a subset of 620 cases in validation set B, three other random dermatologists from the same pool (who had not seen the case previously) graded the cases independently, following the exact same labeling procedure as before. Reference standard differential diagnoses from the two panels of three dermatologists had an AO of 0.61 and an agreement of 0.71 for the primary diagnosis (compared to an AO of 0.51 and an agreement of 0.58 between two random individual dermatologists, one per panel), when considered in the space of 421 mapped conditions. Within the space of 27 conditions handled by DLS, the two panels had an AO of 0.67 and an agreement of 0.73 (compared to an AO of 0.56 and an agreement of 0.62 between two random individual dermatologists, one per panel).

# Supplementary Figures

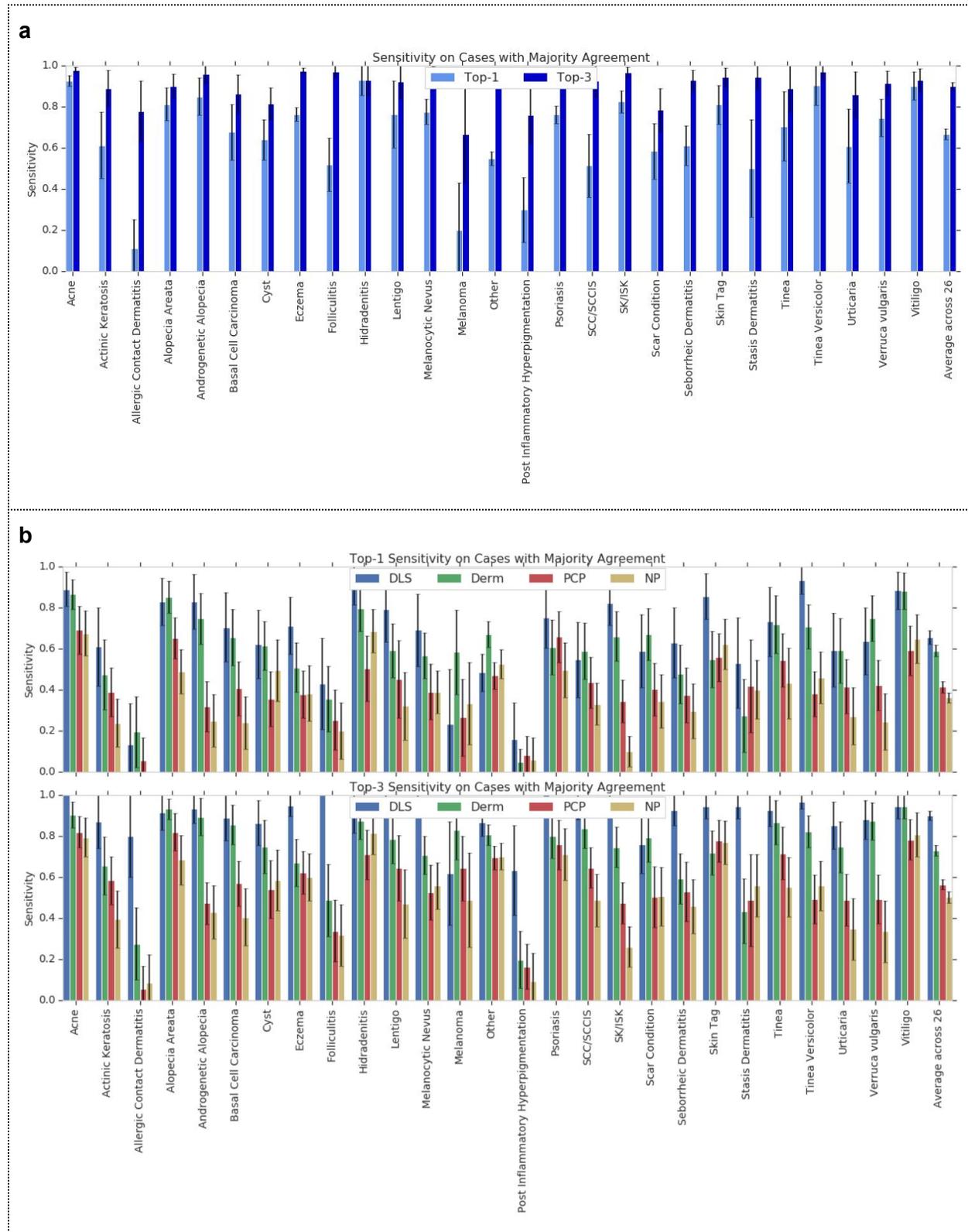

**Supplementary Fig. 1 | Performance of the deep learning system (DLS) and clinicians in cases where at least two out of the three dermatologists determining the reference standard agreed on the primary diagnosis, broken down for each of the 26 categories of skin conditions. a**, Top-1 and top-3 sensitivity of the DLS on validation set A. **b**, Top-1 and top-3 sensitivity of the DLS and three types of clinicians: dermatologists (Derm), primary care physicians (PCP), and nurse practitioners (NP). The number of cases per condition are presented in Supplementary Table 3. The rightmost columns indicate the average sensitivity for the 26 conditions. Error bars indicate 95% confidence intervals.

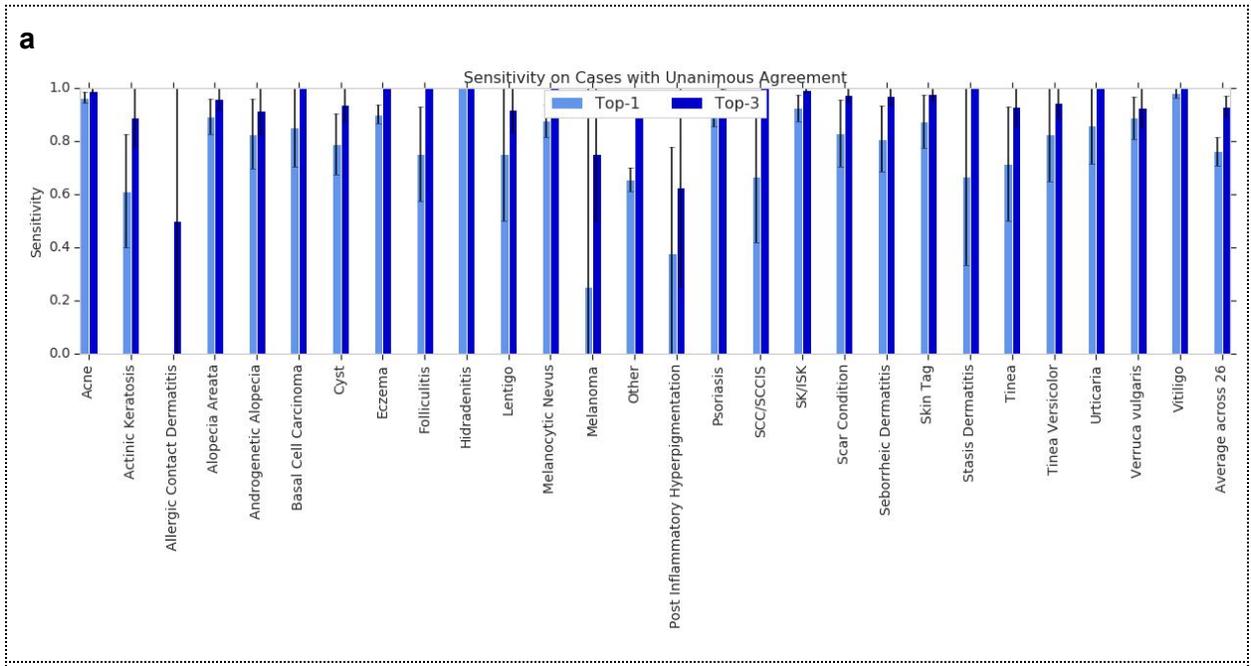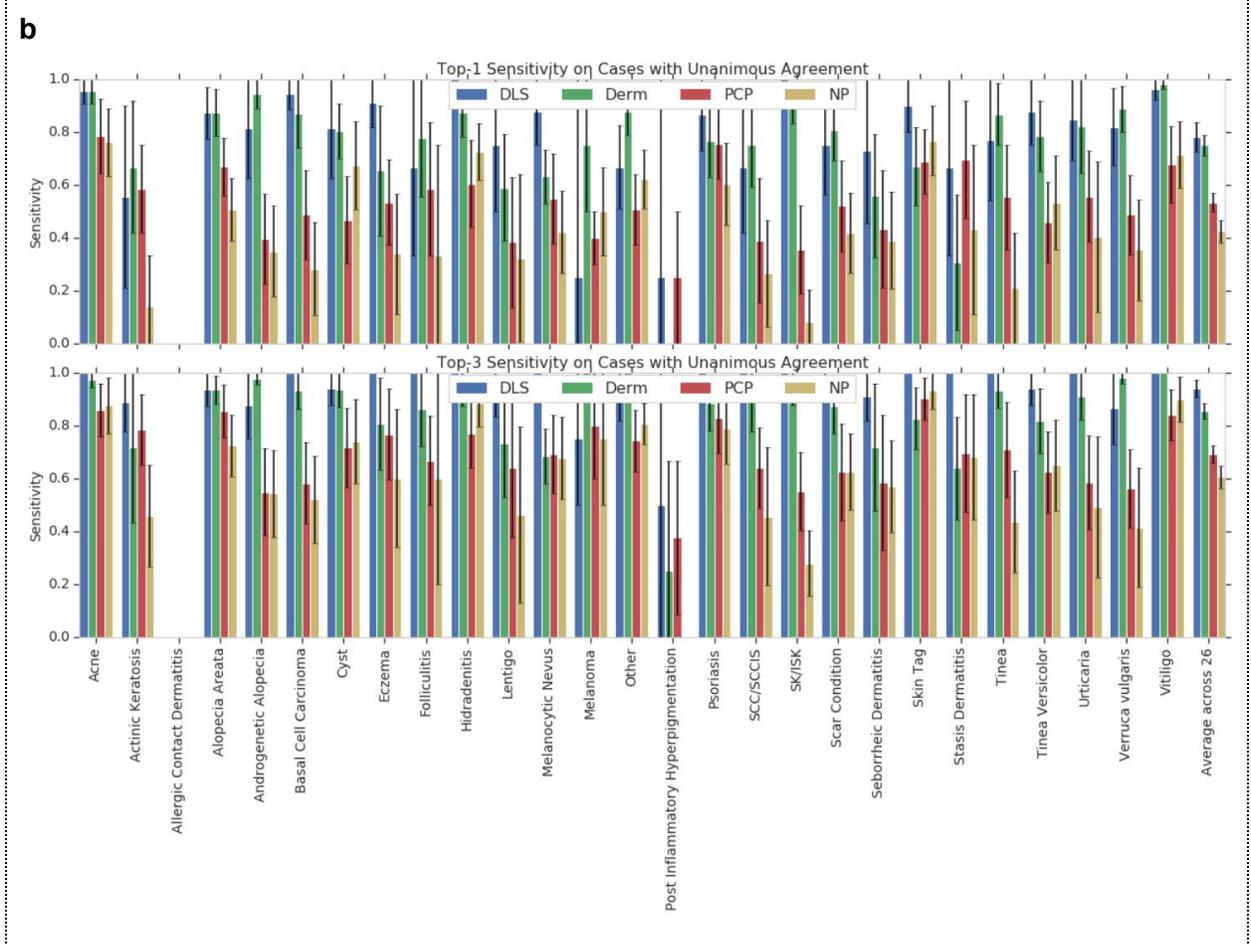

**Supplementary Fig. 2 | Performance of the deep learning system (DLS) and clinicians in cases where all three dermatologists determining the reference standard agreed on the primary diagnosis, broken down for each of the 26 categories of skin conditions. a**, Top-1 and top-3 sensitivity of the DLS on validation set A. **b**, Top-1 and top-3 sensitivity of the DLS and three types of clinicians: dermatologists (Derm), primary care physicians (PCP), and nurse practitioners (NP). The number of cases per condition are presented in Supplementary Table 3. The empty bars for the DLS and all clinicians for allergic contact dermatitis are due to the lack of cases that achieved full consensus for that condition. The rightmost columns indicate the average sensitivity for the 26 conditions. Error bars indicate 95% confidence intervals.

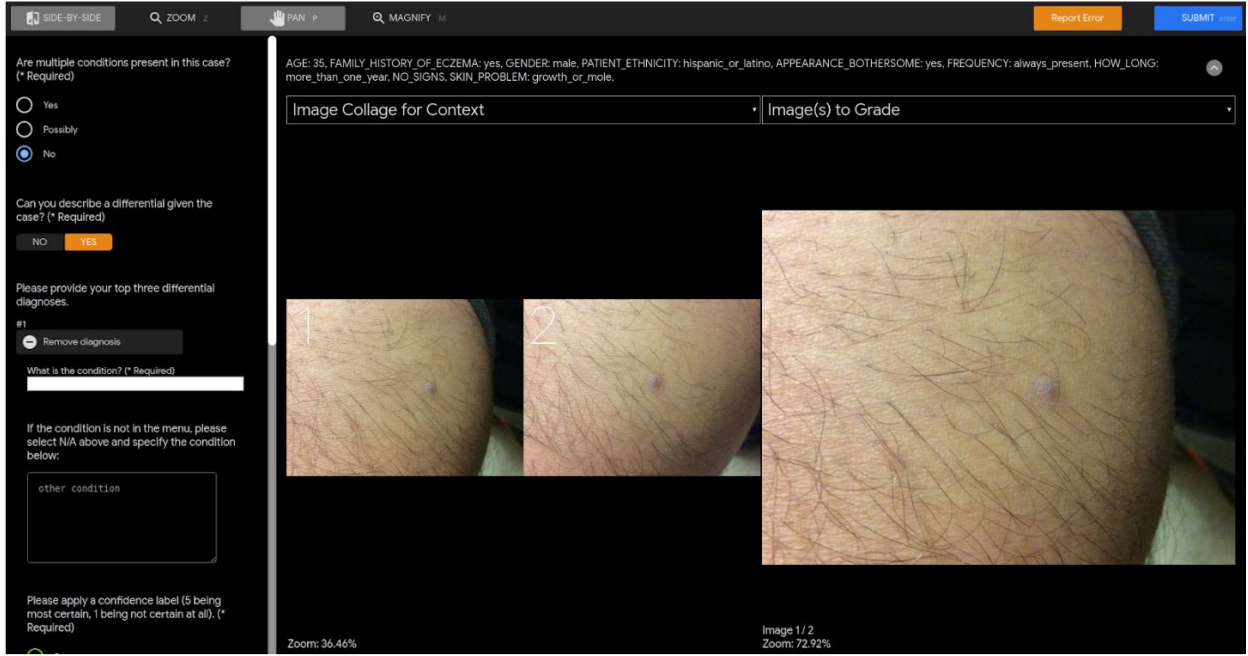

**Supplementary Fig. 3 | Labeling tool interface.** Questions prompts (Supplementary Table 5) are displayed in the left panel, whereas clinical metadata (Supplementary Table 1) are shown in the top right panel and images (up to six per case) are shown in the bottom right panel. Any image could be panned, zoomed, and magnified for closer review.

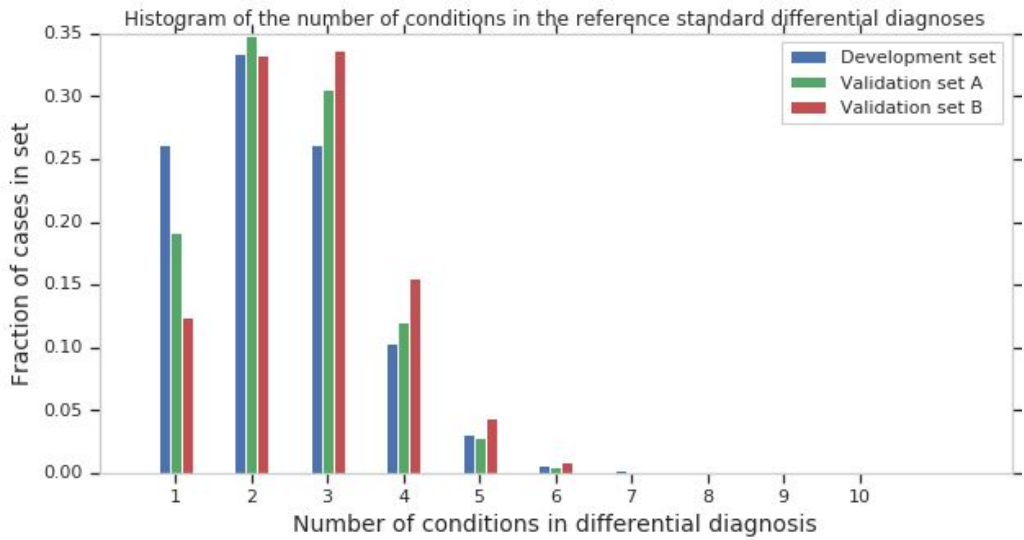

**Supplementary Fig. 4 | Histogram of the number of conditions in the reference standard differential diagnoses.** Within each set (development and validation set A and B), the differential diagnoses has a median length of 2 and a 75$^{th}$ percentile length of 3. The 25$^{th}$ percentile length was slightly different at 1, 2, 2 for the development set, validation set A, and validation set B, respectively.

a

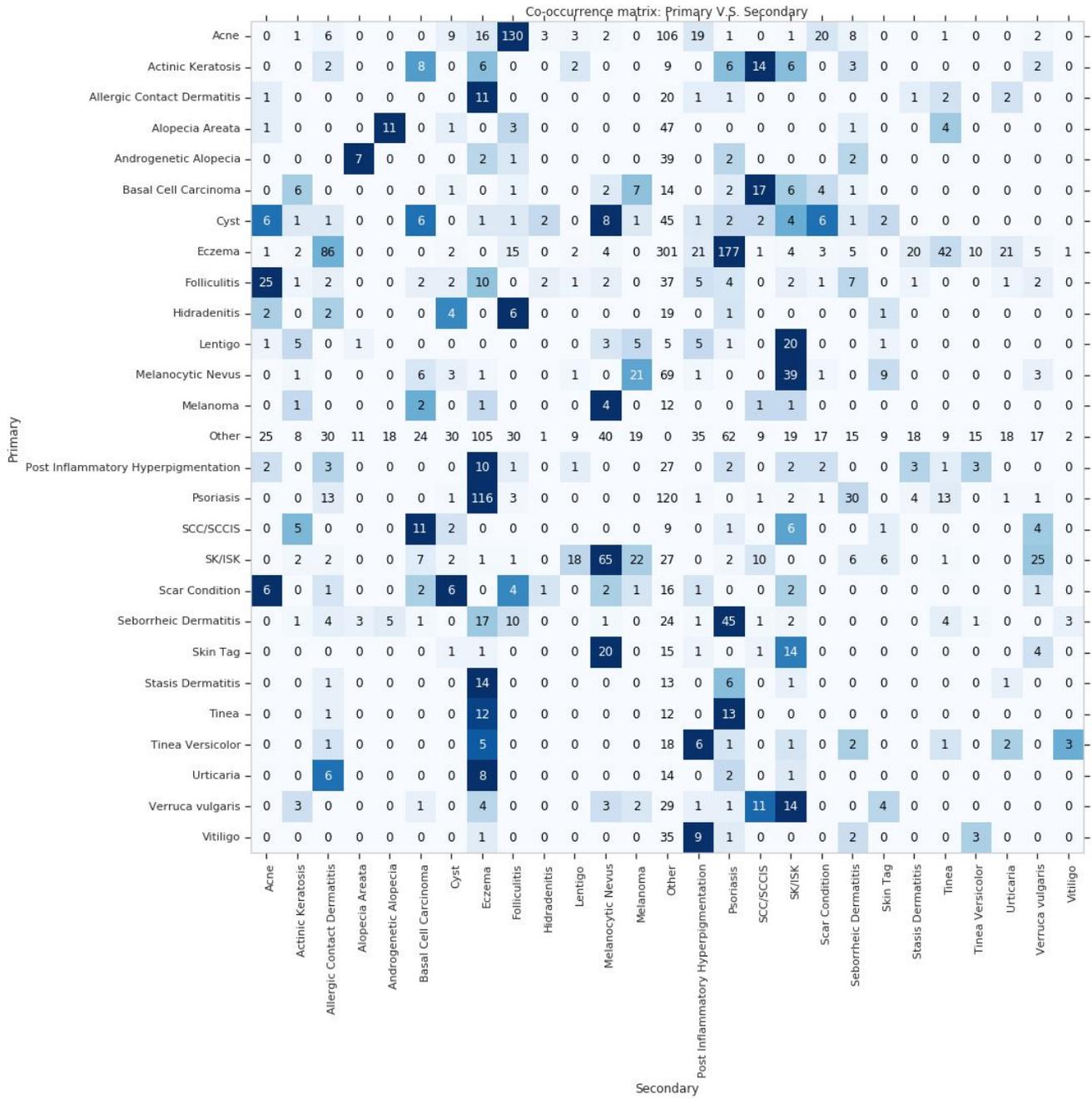

b
51

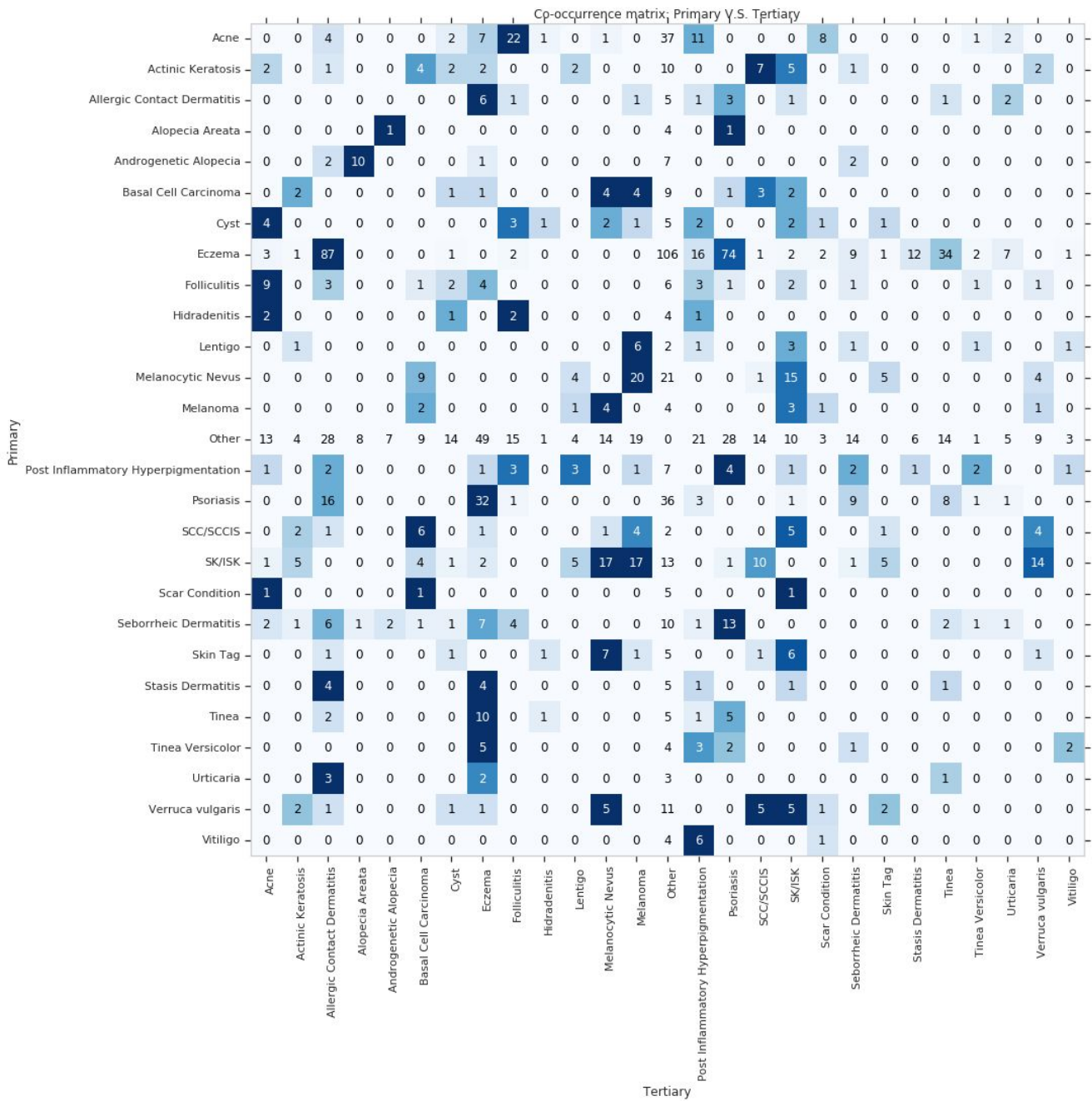

**Supplementary Fig. 5 | Relationship between the primary, secondary, and tertiary diagnoses in the reference standard differential diagnosis in validation set A. a**, Co-occurrence matrix representing the secondary diagnosis for each primary diagnosis. **b**, Co-occurrence matrix representing the tertiary diagnosis for each primary diagnosis. Eczema and psoriasis frequently appear together in the differential, and the same applies for other pairs like eczema and tinea, melanocytic nevus and Seborrheic keratosis / irritated seborrheic

keratosis (SK/ISK), and acne and folliculitis. These pairs share visual similarities which can account for their co-occurrence in the same differential diagnosis.

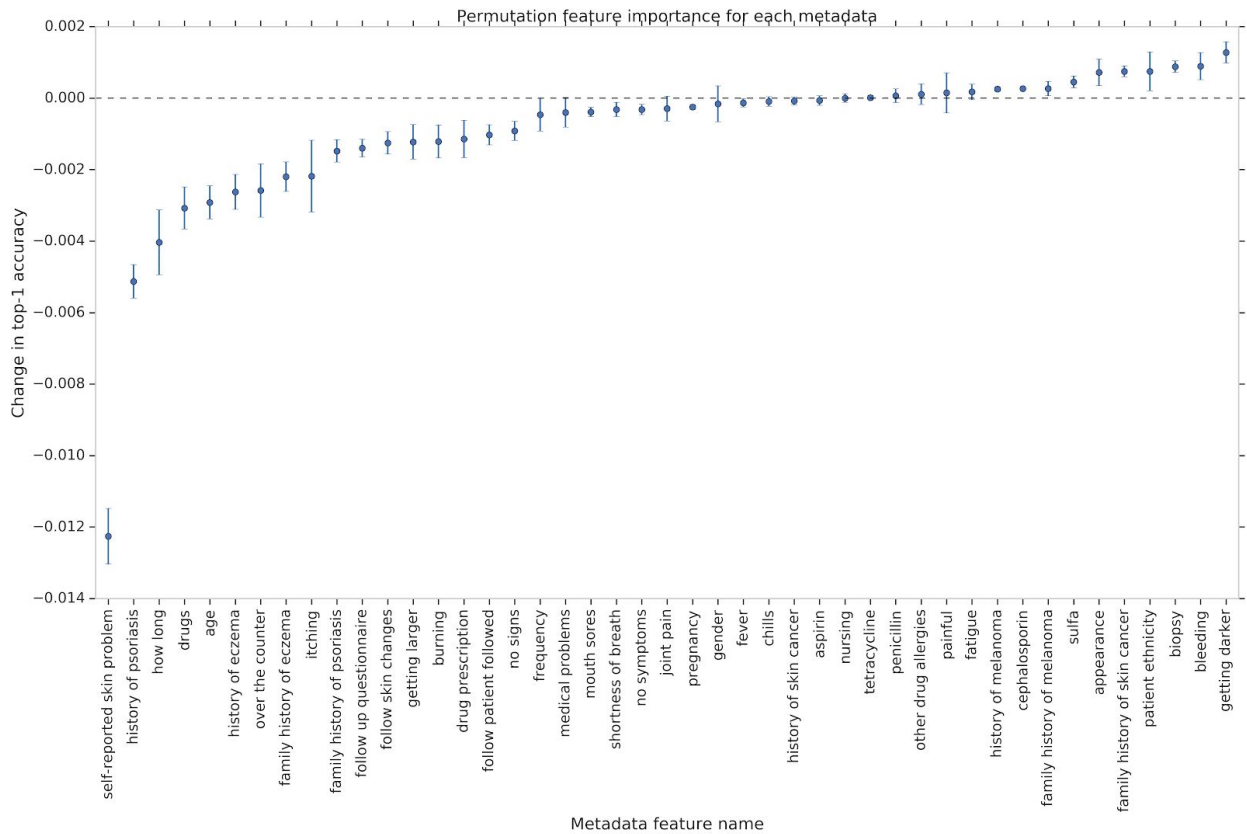

**Supplementary Fig. 6 | Importance of each individual clinical metadata to the deep learning system (DLS).** For each clinical metadata, its values are permuted across validation set A examples, and the effect of this permutation on the top-1 accuracy using the same trained DLS are shown.

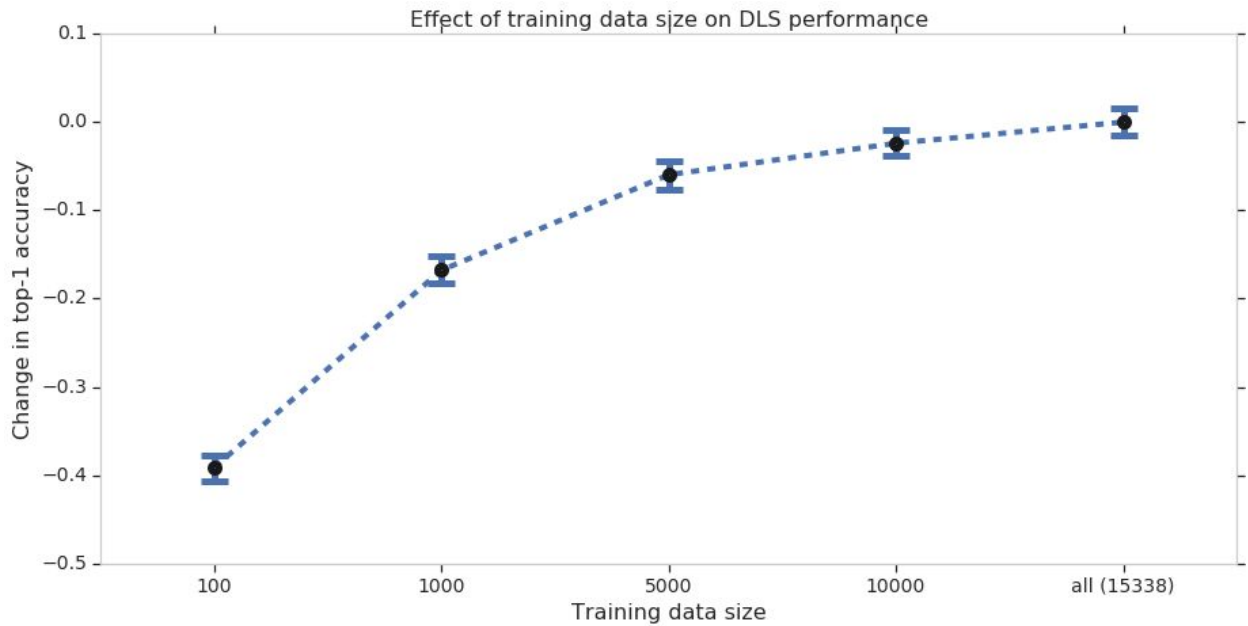

**Supplementary Fig. 7 | Effect of training dataset size on the performance of the deep learning system (DLS).** For each experiment, a random subset of the cases was used for training. This DLS was then evaluated on the validation set A and its change in the top-1 accuracy relative to the original DLS (trained with all available training data) is shown. Error bars indicate 95% confidence intervals across all cases in validation set A.

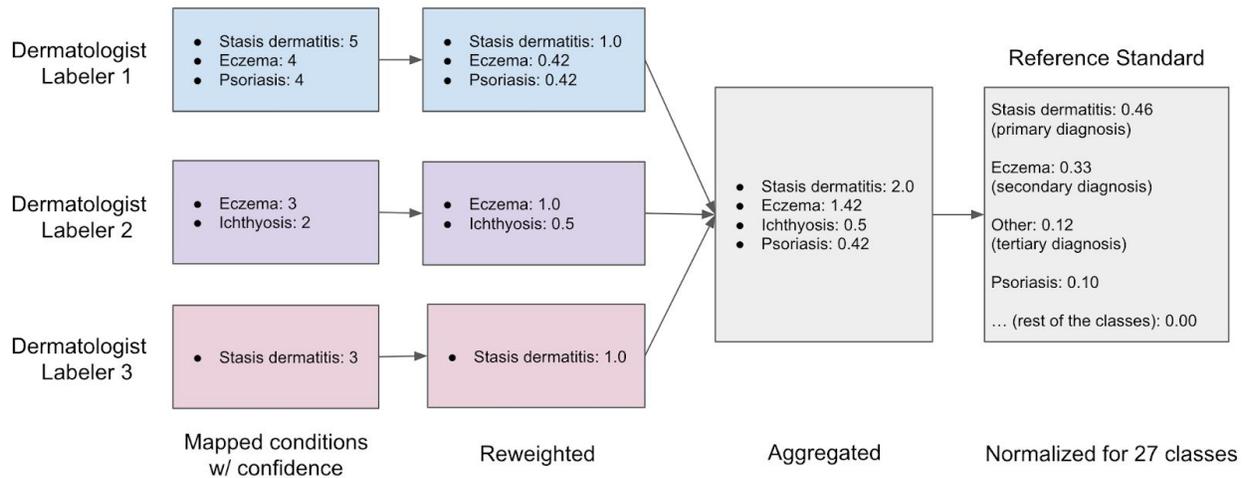

**Supplementary Fig. 8 | Illustration of the establishment of reference standard differential diagnosis.** In this example, each of the three dermatologists reviewed the case independently and provided a list of diagnoses, each with a confidence value ranging from 1 to 5. Weight for each diagnosis (mapped to the 421 list) was determined as the inverse of the rank within each labeler. For the first labeler, since there was a tie between eczema and psoriasis, weights for those were adjusted to evenly distribute between these two ((½ + ⅓ ) / 2 = 0.42). Answers from different labelers were then aggregated by summing up the weights, before limiting the skin condition classes to 27 and normalizing their weights to sum to 1.

| | | | | | | | | |
|---|---|---|---|---|---|---|---|---|
| a | | | | | | | 59 year old (y.o.) Male, White<br>Self-reported skin problem: Rash<br>Duration: More than one year, always present<br>Symptoms: Bothersome in appearance, bleeding, increasing in size<br>Review of system (ROS): No fever/chills (F/C), fatigue, joint pain, mouth sores, or shortness of breath<br>Drugs: Treated by prescription (Rx) or over-the-counter (OTC)<br>Medical history: No history of skin cancer, melanoma, eczema, psoriasis, or biopsy<br>Family history: Psoriasis<br>Drug allergies: None<br>Medication: None<br>Follow-up case?: No | |

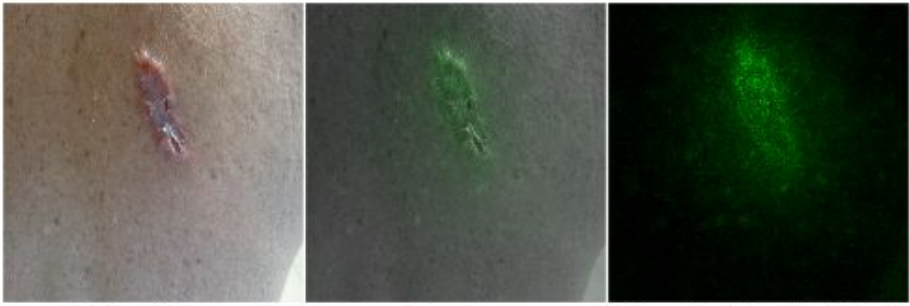

Original Image — Overlay — Integrated Gradient Mask

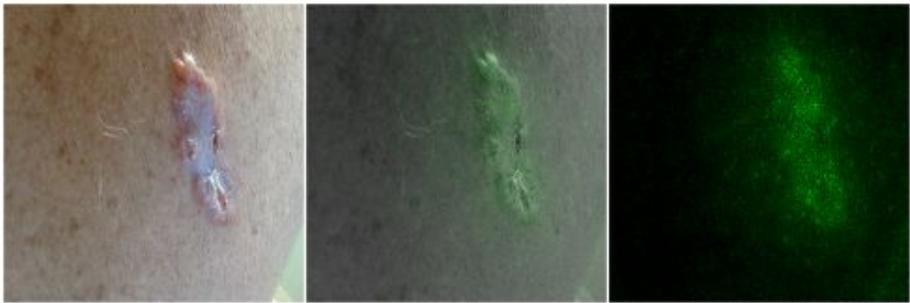

Original Image — Overlay — Integrated Gradient Mask

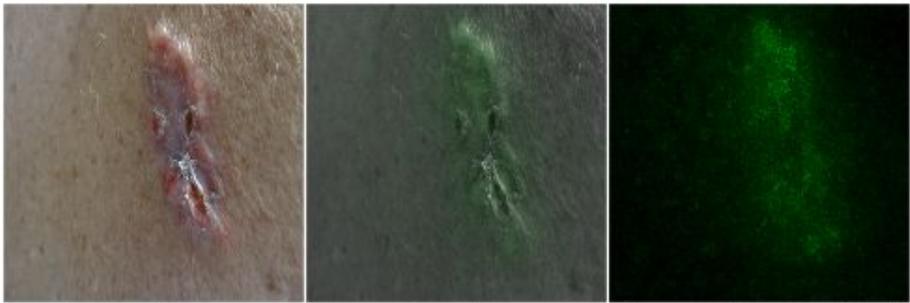

Original Image — Overlay — Integrated Gradient Mask

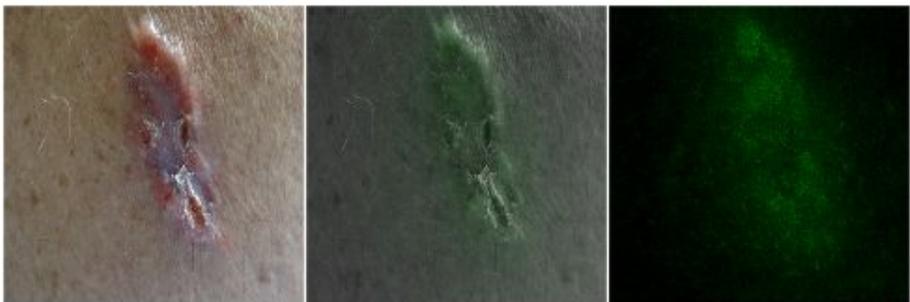

Original Image — Overlay — Integrated Gradient Mask

| Reference standard | DLS (top 3) | DLS (growth subgroup) | NP (missed) | NP (missed) | PCP (tied 1st diagnosis) | PCP (missed) | Derm | Derm |
|---|---|---|---|---|---|---|---|---|
| BCC; SCC/SCCIS; Scar condition | BCC: 0.84; Scar condition: 0.06; SCC/SCCIS: 0.05 | Malignant: 1.0; Benign: 0.0 | Other (hypertrophic skin); Scar condition | AK; Other (skin lesion); Psoriasis | BCC / SCC/SCCIS; Melanoma | Psoriasis | BCC | BCC |

**b**

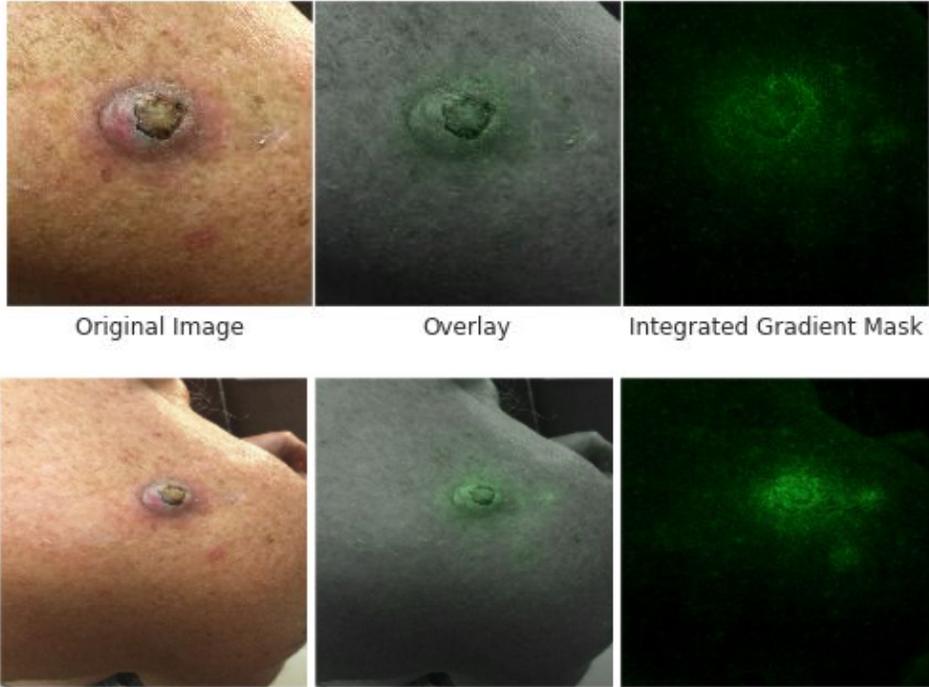

63 y.o. Male
Self-reported skin problem: Growth or Mole
Duration: Three to twelve months, always present
Symptoms: Increasing in size, itching, burning, painful
ROS: No F/C, fatigue, joint pain, mouth sores, or shortness of breath
Drugs: Treated by Rx or OTC
Medical history: No history of skin cancer, melanoma, eczema, psoriasis, or biopsy
Family history: Skin cancer
Drug allergies: None
Medication: None
Follow-up case?: No

| Reference standard | DLS (top 3) | DLS (growth subgroup) | NP (2nd diagnosis) | NP (tied 1st diagnosis) | PCP (missed) | PCP (missed) | Derm | Derm |
|---|---|---|---|---|---|---|---|---|
| SCC/SCCIS; BCC | SCC/SCCIS: 0.74; BCC: 0.19; Actinic keratosis: 0.04 | Malignant: 0.94; Benign: 0.06 | BCC; SCC/SCCIS; Melanoma | Other (skin lesion) / SCC/SCCIS; BCC | Cannot diagnose | Other (pyoderma) | SCC/SCCIS; BCC | SCC/SCCIS |

**c**

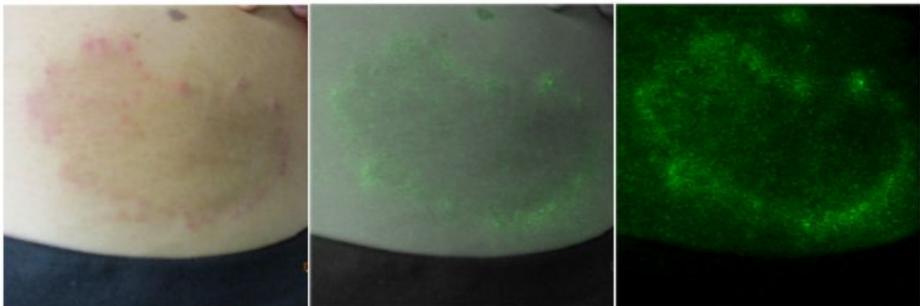

61 y.o. Female, Hispanic or Latino
Self-reported skin problem: Rash
Duration: One to four weeks, always present
Symptoms: Itching
ROS: No F/C, fatigue, joint pain, mouth sores, or shortness of breath
Drugs: Treated by Rx or OTC
Medical history: No history of skin cancer, melanoma, eczema, psoriasis, or biopsy
Family history: Skin cancer
Drug allergies: None
Medication: None

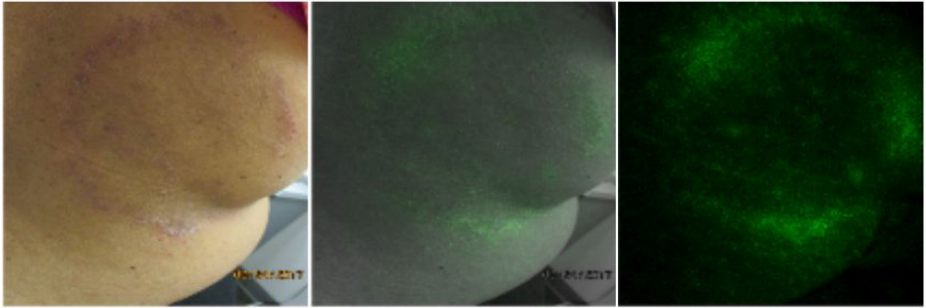

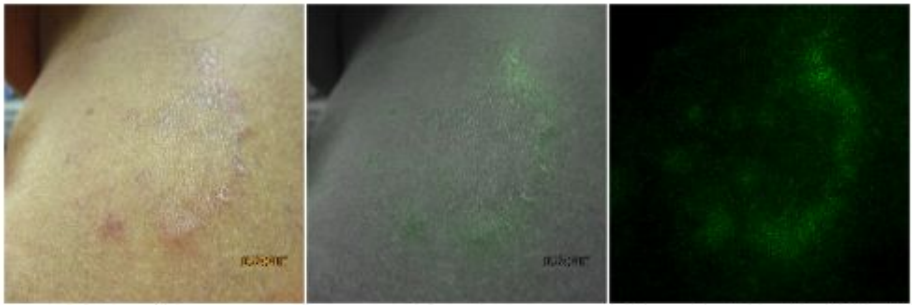

| Reference standard | DLS (top 3) | DLS (erythematosquamous and papulosquamous subgroup) | NP (missed) | NP (missed) | PCP (tied 1st diagnosis) | PCP (missed) | Derm | Derm |
|---|---|---|---|---|---|---|---|---|
| Tinea | Tinea: 0.95; Other: 0.03; Eczema: 0.02 | Infectious: 0.98; Non-infectious: 0.02 | Eczema / Other (Chronic contact dermatitis); Psoriasis | Other (Generalized granuloma annulare) | Other (Granuloma annulare) / Tinea | Eczema | Tinea; Other (Granuloma annulare) | Tinea |

**d**

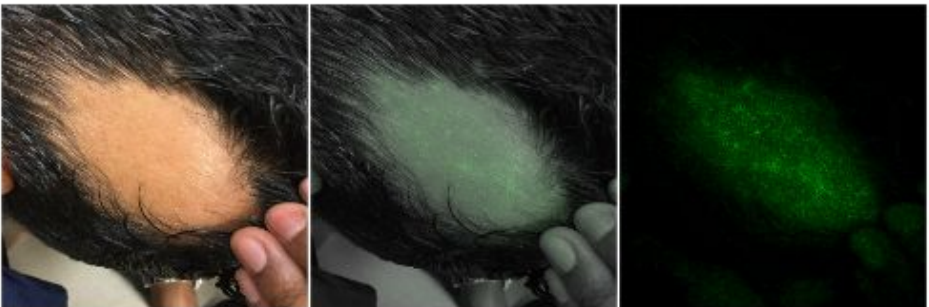

29 y.o. Male, Native Hawaiian or Pacific Islander
Duration: Three to twelve months, always present
Symptoms: Bothersome in appearance, increasing in size, itching
ROS: No F/C, fatigue, joint pain, mouth sores, or shortness of breath
Drugs: Treated by Rx or OTC
Medical history: No history of skin cancer, melanoma, eczema, psoriasis, or biopsy
Family history: Skin cancer
Drug allergies: None
Medication: None
Follow-up case?: Yes

Follow-up case?: No

| | | | | | | | | | |
|---|---|---|---|---|---|---|---|---|---|
| 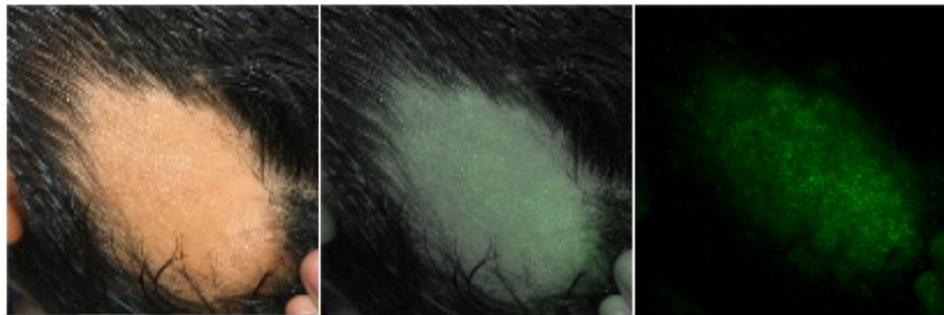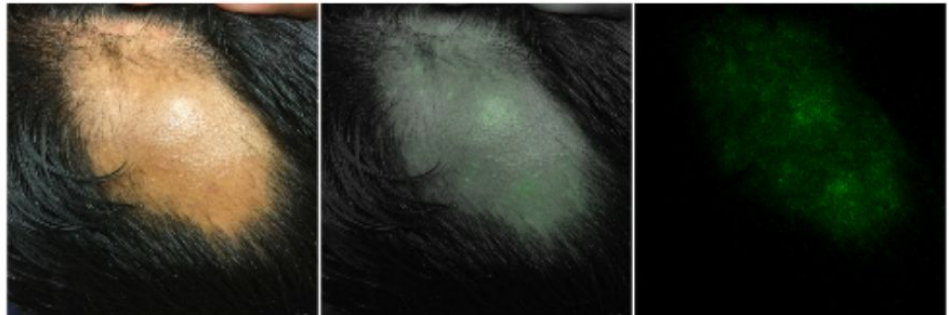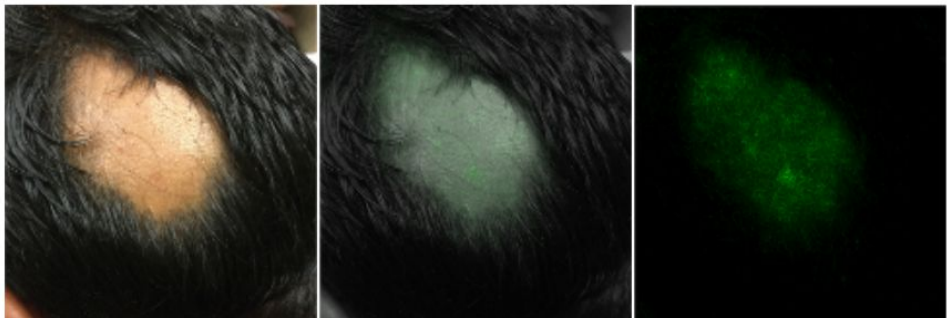 | | | | | | | Patient adhered to treatments: Yes<br>Condition after treatments: Not changed | | |
| **Reference standard** | **DLS (top 3)** | **DLS (hair loss subgroup)** | **NP (3rd diagnosis)** | **NP (missed)** | **PCP (2nd diagnosis)** | **PCP** | **Derm** | **Derm** |
| AA | AA: 0.89;<br>Other: 0.05;<br>AGA: 0.03 | AA: 0.97;<br>AGA: 0.03 | AGA;<br>Other (Alopecia localis);<br>AA | AGA | AGA;<br>AA | AA | AA | AA;<br>Other (trichotillomania) |

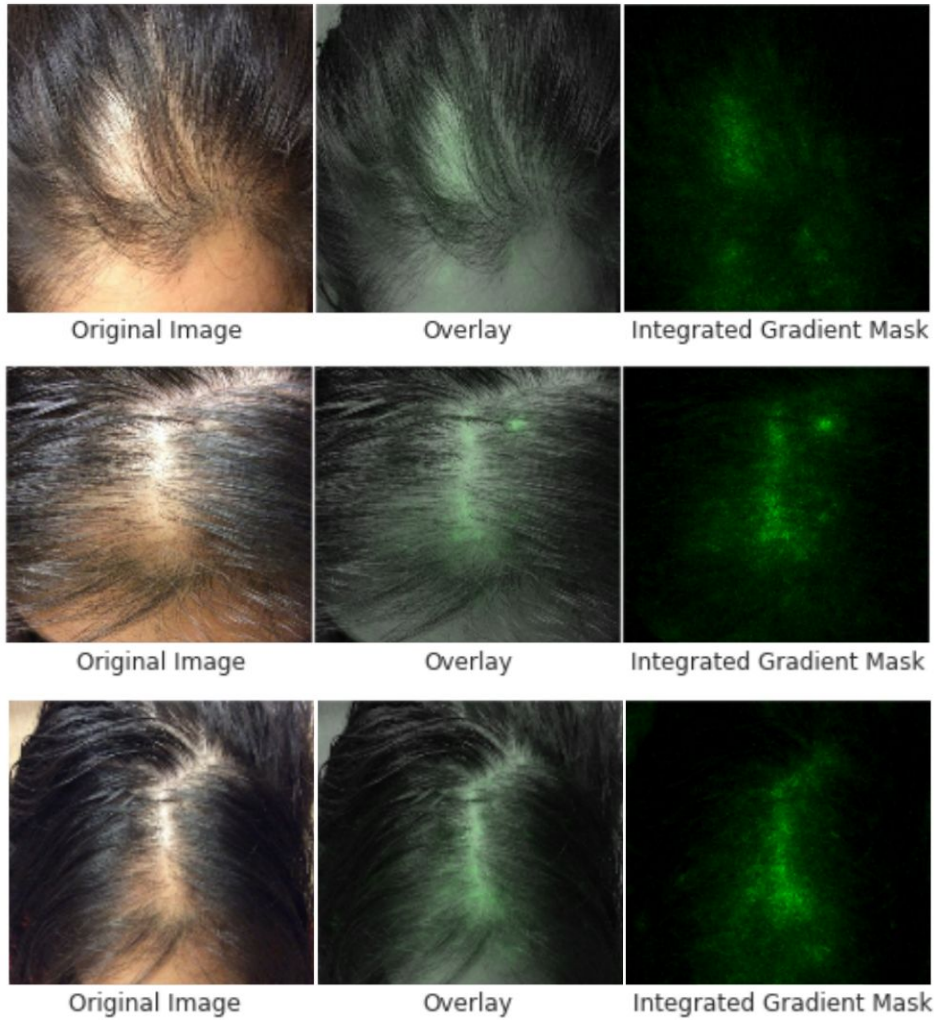

| Reference standard | DLS (top 3) | DLS (hair loss subgroup) | NP (missed, non-specific) | NP (missed, non-specific) | PCP (tied 2ᵈ diagnosis) | PCP (missed) | Derm | Derm |
|---|---|---|---|---|---|---|---|---|
| AGA; AA; Other | AGA: 0.65; Other: 0.25; Seborrheic Dermatitis: 0.05 | AGA: 0.96; AA: 0.04 | Other (Diffuse alopecia) | Other (Alopecia) | Other (Telogen effluvium); AA / AGA | AA | AGA; Other (Drug-related alopecia) | AGA |

26 y.o. Female, Hispanic or Latino
Self-reported skin problem: Hair loss
Duration: More than five years, always present
Symptoms: Bothersome in appearance, itching
ROS: No F/C, fatigue, joint pain, mouth sores, or shortness of breath
Drugs: Has not been treated by Rx or OTC
Medical history: No history of skin cancer, melanoma, eczema, psoriasis, or biopsy
Family history: Skin cancer, eczema
Drug allergies: None
Medication: OTC
Follow-up case?: No

**Supplementary Fig. 9 | All the images and metadata for examples shown in Fig. 3.**
Abbreviations for diagnoses follow those from Fig. 3: basal cell carcinoma (BCC), squamous cell carcinoma (SCC/SCCIS), Alopecia Areata (AA), and Androgenetic Alopecia (AGA). Some images were cropped to zoom in on the condition for clarity.

## Supplementary Tables

**Supplementary Table 1 | Clinical metadata used in this study**

| Name | Description | Possible values |
|---|---|---|
| **Self-reported demographic information** | | |
| Age | The age of the patient in years, at the time the case was submitted. | A float value ranging from 18 to 90. Values larger than 90 are capped at 90. |
| Sex | The sex of the patient. | One of: [Female | Male | Other | Unknown] |
| Race and ethnicity | The race/ethnicity of the patient. | One of: [American Indian or Alaska Native | Asian | Black or African American | Hispanic or Latino | Native Hawaiian or Pacific Islander | White | Neither Hispanic Nor Latino | Not specified | Unknown] |
| **Disease state** | | |
| Self-reported skin problem | The high level skin problem the patient is seeking help for. | One of: [Acne | Growth or mole | Hair loss | Hair or nail problem | Hair problem | Nail problem | Pigmentary problem | Rash | Other | Unknown] |
| Symptoms | Any symptoms perceived by the patient. | A list of 8 symptoms (bothersome in appearance, bleeding, increasing in size, darkening, itching, burning, painful, none of the above) with each symptom being one of: [Yes | No | Unknown]. |
| Signs | Any medical signs perceived by the patient. | A list of 7 signs (fever, chills, fatigue, joint pain, mouth sores, shortness of breath, none of the above) with each sign being one of [Yes | No | Unknown]. |
| Duration | The time that the skin problem has persisted. | One of: [One day | Less than one week | One week | Two weeks | One to four weeks | One month | One to three months | Three months | Three to twelve months | Six months | One year | More than one year | More than five years | Since childhood | Since birth | Unknown] |
| Frequency | Frequency of occurrence of the skin problem. | One of: [Always present | Comes and goes | Unknown] |
| **Medical history** | | |

| Personal history | Personal medical history. | A list of four aspects of the personal history (skin cancer, melanoma, eczema, psoriasis) with each being one of [Yes \| No \| Unknown]. |
|---|---|---|
| Family history | Family medical history. | A list of four aspects of the family history (skin cancer, melanoma, eczema, psoriasis) with each being one of [Yes \| No \| Unknown]. |
| **Patient state** | | |
| Allergy | Medications the patient is allergic to. | A list of 6 allergies (penicillin, cephalosporin, sulfa, tetracycline, aspirin, other) with each being one of [Yes \| No \| Unknown]. |
| Drug | If the patient is currently taking any medications. | One of [Yes \| No \| Unknown]. |
| Pregnancy | If the patient is pregnant. | One of [Yes \| No \| Unknown]. |
| Nursing | If the patient is nursing. | One of [Yes \| No \| Unknown]. |
| Medical problem | Whether the patient currently has any medical problems. | One of [Yes \| No \| Unknown]. |
| **Previous treatment state** | | |
| Follow-up case | If this is a follow up case. | One of [Yes \| No \| Unknown]. |
| Biopsy | If there has been a previous biopsy. | One of [Yes \| No \| Unknown]. |
| Past medication | Whether the patient used medications for the skin problem. | A list of two past medications (prescription drugs, over the counter drugs) with each being one of [Yes \| No \| Unknown]. |
| Patient adhered to treatments | Whether the patient is following the treatment If the patient received treatment before. | One of: [No \| Partially \| Yes \| Unknown] |
| Condition after treatments | Progression of the skin problem If the patient received treatment before. | One of: [Improved \| Not changed \| Worsened \| Unknown] |

**Supplementary Table 2 | Top-1 and top-3 diagnostic accuracy for the three types of clinicians (dermatologists, Derm; primary care physicians, PCP; and nurse practitioners, NP) on validation set B.** Each clinician graded approximately one-third of the cases (number of cases graded: median = 321, range 320-322). For each clinician, performance of the deep learning system (DLS) is also reported on the same cases graded by that clinician (shaded in gray). Bold indicates the higher of the two: clinician or DLS based on each evaluation metric. In particular, Accuracy$_{any}$ measures the agreement of the top-1 and top-3 diagnoses with *any* of the panel of three dermatologists comprising the reference standard.

| Clinician Type / ID | Top 1 | | | | Top 3 | | | | Average Overlap (AO) | |
|---|---|---|---|---|---|---|---|---|---|---|
| | Accuracy | | Accuracy$_{any}$ | | Accuracy | | Accuracy$_{any}$ | | | |
| | Clinician | DLS | Clinician | DLS | Clinician | DLS | Clinician | DLS | Clinician | DLS |
| Derm 1 | 0.58 | **0.67** | 0.72 | **0.79** | 0.68 | **0.92** | 0.79 | **0.97** | 0.54 | **0.64** |
| Derm 2 | 0.65 | **0.66** | **0.83** | 0.75 | 0.80 | **0.86** | 0.92 | **0.93** | **0.62** | 0.61 |
| Derm 3 | 0.65 | **0.68** | 0.81 | **0.85** | 0.78 | **0.91** | 0.91 | **0.97** | 0.59 | **0.64** |
| Derm 4 | 0.61 | **0.66** | 0.75 | **0.80** | 0.69 | **0.91** | 0.82 | **0.96** | 0.54 | **0.63** |
| Derm 5 | 0.64 | **0.68** | 0.74 | **0.80** | 0.80 | **0.90** | 0.88 | **0.94** | 0.60 | **0.62** |
| Derm 6 | 0.64 | **0.67** | 0.76 | **0.79** | 0.73 | **0.89** | 0.84 | **0.97** | 0.56 | **0.65** |
| PCP 1 | 0.43 | **0.70** | 0.54 | **0.81** | 0.66 | **0.92** | 0.80 | **0.97** | 0.46 | **0.67** |
| PCP 2 | 0.50 | **0.66** | 0.63 | **0.76** | 0.74 | **0.86** | 0.86 | **0.93** | 0.54 | **0.61** |
| PCP 3 | 0.44 | **0.65** | 0.61 | **0.81** | 0.53 | **0.91** | 0.70 | **0.97** | 0.43 | **0.63** |
| PCP 4 | 0.48 | **0.68** | 0.62 | **0.80** | 0.50 | **0.91** | 0.63 | **0.96** | 0.42 | **0.64** |
| PCP 5 | 0.44 | **0.66** | 0.58 | **0.78** | 0.51 | **0.89** | 0.65 | **0.95** | 0.43 | **0.63** |
| PCP 6 | 0.39 | **0.66** | 0.54 | **0.80** | 0.65 | **0.89** | 0.81 | **0.96** | 0.47 | **0.63** |
| NP 1 | 0.42 | **0.70** | 0.58 | **0.81** | 0.53 | **0.91** | 0.67 | **0.96** | 0.44 | **0.65** |
| NP 2 | 0.36 | **0.63** | 0.45 | **0.73** | 0.55 | **0.87** | 0.72 | **0.93** | 0.42 | **0.61** |
| NP 3 | 0.42 | **0.68** | 0.56 | **0.84** | 0.58 | **0.91** | 0.73 | **0.97** | 0.44 | **0.64** |
| NP 4 | 0.38 | **0.67** | 0.55 | **0.78** | 0.39 | **0.90** | 0.56 | **0.94** | 0.36 | **0.64** |
| NP 5 | 0.42 | **0.69** | 0.51 | **0.83** | 0.60 | **0.91** | 0.72 | **0.96** | 0.43 | **0.64** |
| NP 6 | 0.44 | **0.64** | 0.59 | **0.78** | 0.65 | **0.88** | 0.79 | **0.97** | 0.47 | **0.62** |

**Supplementary Table 3 | Number of cases per category of skin condition, filtered by different levels of agreement on the primary diagnosis among dermatologists determining the reference standard.**

| Condition name | Validation set A | | | Validation set B | | |
|---|---|---|---|---|---|---|
| | No. of cases | No. of cases with agreement by ≥2 dermatologists (%) | No. of cases with agreement by all 3 dermatologists (%) | No. of cases | No. of cases with agreement by ≥2 dermatologists (%) | No. of cases with agreement by all 3 dermatologists (%) |
| Acne | 428 | 381 (89.0%) | 267 (62.4%) | 47 | 36 (76.6%) | 21 (44.7%) |
| Actinic Keratosis | 57 | 36 (63.2%) | 18 (31.6%) | 40 | 23 (57.5%) | 9 (22.5%) |
| Allergic Contact Dermatitis | 49 | 27 (55.1%) | 2 (4.1%) | 36 | 15 (41.7%) | 0 (0.0%) |
| Alopecia Areata | 98 | 90 (91.8%) | 73 (74.5%) | 39 | 35 (89.7%) | 31 (79.5%) |
| Androgenetic Alopecia | 56 | 46 (82.1%) | 23 (41.1%) | 36 | 29 (80.6%) | 16 (44.4%) |
| Basal Cell Carcinoma | 48 | 43 (89.6%) | 27 (56.2%) | 31 | 27 (87.1%) | 17 (54.8%) |
| Cyst | 97 | 80 (82.5%) | 47 (48.5%) | 37 | 29 (78.4%) | 16 (43.2%) |
| Eczema | 719 | 565 (78.6%) | 229 (31.8%) | 71 | 38 (53.5%) | 11 (15.5%) |
| Folliculitis | 111 | 64 (57.7%) | 20 (18.0%) | 43 | 21 (48.8%) | 6 (14.0%) |
| Hidradenitis | 47 | 41 (87.2%) | 31 (66.0%) | 37 | 32 (86.5%) | 23 (62.2%) |
| Lentigo | 37 | 25 (67.6%) | 12 (32.4%) | 36 | 24 (66.7%) | 12 (33.3%) |
| Melanocytic Nevus | 195 | 168 (86.2%) | 96 (49.2%) | 40 | 29 (72.5%) | 16 (40.0%) |
| Melanoma | 27 | 15 (55.6%) | 4 (14.8%) | 22 | 13 (59.1%) | 4 (18.2%) |
| Post Inflammatory Hyperpigmentation | 66 | 37 (56.1%) | 8 (12.1%) | 38 | 19 (50.0%) | 4 (10.5%) |
| Psoriasis | 365 | 316 (86.6%) | 199 (54.5%) | 49 | 32 (65.3%) | 22 (44.9%) |
| SCC/SCCIS | 38 | 39 (102.6%) | 12 (31.6%) | 35 | 33 (94.3%) | 12 (34.3%) |
| SK/ISK | 224 | 203 (90.6%) | 118 (52.7%) | 44 | 39 (88.6%) | 22 (50.0%) |
| Scar Condition | 69 | 55 (79.7%) | 35 (50.7%) | 38 | 29 (76.3%) | 20 (52.6%) |
| Seborrheic Dermatitis | 112 | 82 (73.2%) | 31 (27.7%) | 43 | 27 (62.8%) | 11 (25.6%) |
| Skin Tag | 73 | 68 (93.2%) | 39 (53.4%) | 35 | 34 (97.1%) | 20 (57.1%) |
| Stasis Dermatitis | 30 | 18 (60.0%) | 6 (20.0%) | 29 | 17 (58.6%) | 6 (20.7%) |
| Tinea | 38 | 27 (71.1%) | 14 (36.8%) | 35 | 26 (74.3%) | 13 (37.1%) |
| Tinea Versicolor | 37 | 31 (83.8%) | 17 (45.9%) | 36 | 30 (83.3%) | 16 (44.4%) |
| Urticaria | 39 | 28 (71.8%) | 14 (35.9%) | 38 | 27 (71.1%) | 13 (34.2%) |
| Verruca vulgaris | 88 | 82 (93.2%) | 53 (60.2%) | 38 | 33 (86.8%) | 22 (57.9%) |
| Vitiligo | 78 | 70 (89.7%) | 51 (65.4%) | 38 | 34 (89.5%) | 25 (65.8%) |
| Other | 915 | 849 (92.8%) | 390 (42.6%) | 142 | 110 (77.5%) | 33 (23.2%) |

**Supplementary Table 4 | Top-1 and top-3 sensitivity averaged across all the skin conditions categories, and with different exclusions.** Allergic Contact Dermatitis (ACD) and Post-inflammatory Hyperpigmentation (PIH) are included in this analysis because of the low sensitivity for these conditions by both the deep learning system (DLS) and the three types of clinicians (dermatologists, Derms; primary care physicians, PCPs; and nurse practitioners, NPs) on validation set B. Bold indicates the highest value within each row and each evaluation metric for validation set B.

| Conditions included in the average | Average Top-1 Sensitivity | | | | Average Top-3 Sensitivity | | | |
|---|---|---|---|---|---|---|---|---|
| | DLS | Derm | PCP | NP | DLS | Derm | PCP | NP |
| All 27 conditions | **0.57** | 0.52 | 0.36 | 0.33 | **0.82** | 0.64 | 0.50 | 0.45 |
| 26 conditions (excludes "Other") | **0.57** | 0.51 | 0.35 | 0.32 | **0.82** | 0.64 | 0.49 | 0.44 |
| 25 conditions (excludes ACD and PIH) | **0.60** | 0.55 | 0.38 | 0.35 | **0.84** | 0.68 | 0.53 | 0.49 |
| 24 conditions (excludes ACD, PIH, and "Other") | **0.60** | 0.55 | 0.38 | 0.35 | **0.84** | 0.68 | 0.53 | 0.48 |

**Supplementary Table 5 | Labeling tool prompts and instructions.**

| Question | | Possible answers (underlined), with explanations if applicable |
|---|---|---|
| Are multiple conditions present in this case? | | <u>Yes</u>*: if more than one condition related to this patient's chief complaint is present<br><u>Possibly</u>*: if more than one condition may be present<br><u>No</u>: if there is a single skin condition |
| Can you describe a differential given the case? | | <u>Yes</u>: if one can provide a diagnosis.<br><u>No</u>*: if one cannot provide any diagnosis. This can be due to poor image quality, minimum pathology, insufficient medical information, etc. |
| Please provide your top three differential diagnosis: | What is the condition? | <u>SNOMED texts synonyms</u>: an autocomplete menu that contains all synonyms for SNOMED entries pertaining to cutaneous disease is available to select from. If there are several variations of the condition, use the most specific condition that applies to the case. If none found, then:<br><u>Free text</u>: an additional text field is provided for labelers to enter any free-form text. |
| | Confidence of diagnosis | <u>5</u>: most certain about the condition.<br><u>4</u>:<br><u>3</u>:<br><u>2</u>:<br><u>1</u>: least certain about the condition. |

* If these answers are selected, the remaining questions are skipped.

**Supplementary Table 6 | Full list of 421 skin conditions that answers from dermatologists, PCPs, and NPs were mapped to.** The top 26 conditions on which the DLS was trained and evaluated on are highlighted in bold. The remaining 395 conditions (in aggregate comprising roughly 20% of the cases in this dataset) were mapped to "Other" for the purposes of this study.

| A-C | D-H | I-M | N-P | R-Z |
|---|---|---|---|---|
| Abscess | Deep fungal infection | Ichthyosis | Nail dystrophy due to trauma | RMSF - Rocky Mountain spotted fever |
| Acanthoma fissuratum | Dental fistula | Idiopathic exfoliative cheilitis | Nasal polyp | Radiation dermatitis |
| Acanthosis nigricans | Dermatitis herpetiformis | Idiopathic guttate hypomelanosis | Nasolabial dyssebacia | Raynaud's phenomenon |
| Accessory nipple | Dermatitis of anogenital region | IgA pemphigus | Necrobiosis lipoidica | Relapsing polychondritis |
| **Acne** | Dermatofibroma | Impetigo | Necrolytic acral erythema | Remove from labeling tool |
| Acne keloidalis | Dermatofibrosarcoma protuberans | Incontinentia pigmenti | Necrotizing fasciitis | Retention hyperkeratosis |
| Acquired digital fibrokeratoma | Dermatomyositis | Induced hypopigmentation | Neuralgia paresthetica | Reticular erythematous mucinosis |
| Acral keratosis | Dermatosis caused by lice | Infected eczema | Neutrophilic eccrine hidradenitis | Reticulate erythematous mucinosis |
| Acral peeling skin syndrome | Dermoid cyst of skin | Infected skin ulcer | Nevus anemicus | Reticulohistiocytosis |
| Acrocyanosis | Desmoplastic trichoepithelioma | Inflammatory linear verrucous epidermal nevus | Nevus comedonicus | Rheumatoid nodule |
| Acrodermatitis atrophicans chronica | Diabetic dermopathy | Inflicted skin lesions | Nevus depigmentosus | Rhytides |
| Acropustulosis of infancy | Diabetic ulcer | Ingrown hair | Nevus lipomatosus cutaneous superficialis | Rosacea |
| **Actinic Keratosis** | Digital Myxoid Cyst | Injection site disorder | Nevus of Ito | **SCC/SCCIS** |
| Actinic granuloma | Digital mucous cyst | Insect Bite | Nevus of Ota | SJS/TEN |
| Acute generalised exanthematous pustulosis | Dissecting cellulitis of scalp | Interstitial granulomatous dermatitis | Nevus sebaceous | **SK/ISK** |
| Adnexal neoplasm | Dowling-degos syndrome | Intertrigo | Nevus spilus | Scabies |
| Adult onset still disease | Drug Rash | Inverted follicular keratosis | Nodular vasculitis | **Scar Condition** |
| Albinism | Eccrine carcinoma of skin | Irritant Contact Dermatitis | Non-melanin pigmentation due to exogenous substance (disorder) | Scleredema |
| **Allergic Contact Dermatitis** | Ecthyma | Juvenile xanthogranuloma | Notalgia paresthetica | Sclerodactyly |
| **Alopecia Areata** | Ecthyma gangrenosum | Kaposi's sarcoma of skin | O/E - ecchymoses present | Sebaceous adenoma of skin |
| Alopecia mucinosa | **Eczema** | Keratoderma | Ochronosis | Sebaceous carcinoma |
| Alopecia neurotica | Edema bulla | Keratolysis exfoliativa | Onychocryptosis | Sebaceous hyperplasia |
| Amyloidosis of skin | Epidermal nevus | Keratosis pilaris | Onychogryphosis | **Seborrheic Dermatitis** |
| Anagen effluvium | Epidermolysis bullosa | Knuckle pads | Onycholysis | **Skin Tag** |
| **Androgenetic Alopecia** | Erosive pustular dermatosis | **Lentigo** | Onychomadesis | Skin and soft tissue atypical mycobacterial infection |
| Anetoderma | Eruptive xanthoma | Leprosy | Onychomalacia | Skin atrophy |
| Angina bullosa hemorrhagica | Erysipelas | Leukemia cutis | Onychomatricoma | Skin changes due to malnutrition |
| Angioedema | Erythema ab igne | Leukocytoclastic Vasculitis | Onychomycosis | Skin lesion in drug addict |
| Angiofibroma | Erythema annulare centrifugum | Leukonychia | Onychopapilloma | Skin striae |
| Angiokeratoma of skin | Erythema dyschromicum perstans | Leukoplakia of skin | Onychorrhexis | Small plaque parapsoriasis |
| Angiolymphoid hyperplasia with eosinophilia | Erythema elevatum diutinum | Lichen Simplex Chronicus | Onychoschizia | Small vessel thrombosis of skin |
| Angiosarcoma of skin | Erythema gyratum repens | Lichen nitidus | Oral fibroma | Sneddon-Wilkinson disease |
| Animal bite - wound | Erythema marginatum | Lichen planopilaris | Osteoarthritis | **Stasis Dermatitis** |
| Apocrine cystadenoma | Erythema migrans | Lichen planus/lichenoid eruption | Osteoma | Subungual fibroma |
| Arsenical keratosis | Erythema multiforme | Lichen sclerosus | Osteoma cutis | Sweet syndrome |
| Arterial ulcer | Erythema nodosum | Lichen spinulosus | Otitis externa | Symmetrical dyschromatosis of extremities |
| Arteriovenous malformation | Erythrasma | Lichen striatus | Paget disease | Syphilis |
| Atrophic glossitis | Erythromelalgia | Lichenoid keratosis | Palisaded neutrophilic granulomatous dermatitis | TMEP - telangiectasia macularis eruptiva perstans |
| Atrophoderma | Erythromelanosis follicularis faciei et colli | Lichenoid myxedema | Palmar pit | Tattoo |
| Atrophoderma vermiculatum | Fat necrosis | Linear IgA disease | Papilloma of skin | Telangiectasia disorder |
| **Atypical Nevus** | Fibrofolliculoma | Lipoatrophy | Parapsoriasis | Telogen effluvium |
| Atypical fibroxanthoma of skin | Flagellate erythema | Lipodermatosclerosis | Paronychia | Thrombophlebitis |
| B-Cell Cutaneous Lymphoma | Flegels disease | Lipoid proteinosis | Pearly penile papules | **Tinea** |
| **Basal Cell Carcinoma** | Flushing | Lipoma | Pemphigoid gestationis | **Tinea Versicolor** |
| Beau's lines | Focal epithelial hyperplasia of skin | Lipsch303274tz ulcer | Pemphigus foliaceus | Torus palatinus |
| Becker's nevus | **Folliculitis** | Livedo reticularis | Pemphigus paraneoplastica | Trachyonychia |
| | Folliculitis decalvans | Livedoid vasculopathy | Pemphigus vulgaris | |
| | | Lobomycosis | Perforating dermatosis | |
| | | Local infection of wound | Perichondritis of auricle | |
| | | Longitudinal melanonychia | | |
| | | Lymphadenopathy | | |
| Benign neoplasm of nail apparatus | Fordyce spots | Lymphangioma | Perioral Dermatitis | Traction alopecia |
| Benign neural tumor | Foreign body | Lymphedema | Periungual fibroma | Traumatic bulla |
| Benign salivary gland tumor | Foreign body reaction of the skin | Lymphomatoid papulosis | Perleche | Traumatic ulcer |
| Blistering distal dactylitis | Fox-Fordyce disease | Madarosis | Phimosis | Triangular alopecia |
| Blue sacral spot | Frontal fibrosing alopecia | Malignant cylindroma | Photodermatitis | Trichostasis spinulosa |
| Bowenoid papulosis | Ganglion cyst | Malignant eccrine spiradenoma | Phrynoderma | Trichotillomania |
| Brachioradial pruritus | Geographic tongue | Mastocytoma | Piezogenic pedal papule | Trigeminal trophic syndrome |
| Breast cancer | Giant cell tumor | Mastocytosis | Pigmented fungiform papillae | Tripe palms |
| Bullosis diabeticorum | Glomus tumour of skin | Median rhomboid glossitis | Pigmented purpuric eruption | Tuberculosis of skin and subcutaneous tissue |
| Bullous Pemphigoid | Gout | Melanin pigmentation due to exogenous substance | Pilomatricoma | Ulceration in Behcet disease |
| Burn of skin | Graft versus host disease | | Pilonidal cyst | |
| Bursitis | Granular parakeratosis | **Melanocytic Nevus** | Pincer nail deformity | **Urticaria** |
| Cafe au lait macule | Granuloma annulare | **Melanoma** | Pinkus tumor | Urticaria multiforme |
| Calcinosis cutis | Granuloma faciale | Melanotic macule | Pitted keratolysis | Varicose veins of lower extremity |
| Calciphylaxis cutis | Granulomatous cheilitis | Melasma | Pityriasis alba | |
| Candida | Grover's disease | Merkel Cell Carcinoma | Pityriasis amiantacea | Venous Stasis Ulcer |
| Canker sore | Hailey Hailey disease | Microcystic adnexal carcinoma | Pityriasis lichenoides | **Verruca vulgaris** |
| Carotene pigmentation of skin | Hair nevus | Milia | Pityriasis rosea | Viral Exanthem |
| Cellulitis | Hair sinus | Miliaria | Pityriasis rotunda | **Vitiligo** |
| Central centrifugal cicatricial alopecia | Hairy tongue | Molluscum Contagiosum | Pityriasis rubra pilaris | Warty dyskeratoma |
| | Half-and-half nail | Morphea/Scleroderma | Pleomorphic fibroma | Wells' syndrome |
| Chancroid | Hand foot and mouth disease | Morsicatio buccarum | Poikiloderma | Wooly hair |
| Chemical leukoderma | Head lice | Mucocele | Polymorphous Light Eruption | Xanthoma |
| Chicken pox exanthem | Hemangioma | Mucocutaneous venous malformation | Porokeratosis | Xerosis |
| Chilblain | Hematoma of skin | | Porphyria cutanea tarda | Yellow nail syndrome |
| Chondrodermatitis nodularis | Hemorrhoid | | **Post Inflammatory Hyperpigmentation** | Zoon's balanitis |
| Cicatricial Pemphigoid | Hemosiderin pigmentation of skin | | | Zosteriform reticulate hyperpigmentation |
| Clavus | Herpes Simplex | | Post-inflammatory hypopigmentation | |
| Clear cell acanthoma | Herpes Zoster | | Pressure ulcer | |
| Clubbing of fingers | **Hidradenitis** | | Pressure-induced dermatosis | |
| Collagenoma | Hirsutism | | Pretibial myxedema | |
| Colloid milium | Hordeolum internum | | Primary cutaneous sarcoma | |
| Comedone | Hyperhidrosis | | Progressive macular hypomelanosis | |
| Condyloma acuminatum | Hypersensitivity | | Prurigo nodularis | |
| Confluent and reticulate papillomatosis | Hypertrichosis | | Pruritic urticarial papules and plaques of pregnancy | |
| Congenital alopecia | | | Pseudocyst of auricle | |
| Connective tissue nevus | | | Pseudolymphoma | |
| Crohn disease of skin | | | Pseudopelade | |
| Cutaneous T Cell Lymphoma | | | **Psoriasis** | |
| Cutaneous capillary malformation | | | Psychogenic alopecia | |
| Cutaneous collagenous vasculopathy | | | Pterygium of nail | |
| Cutaneous larva migrans | | | Puncture wound - injury | |
| Cutaneous leishmaniasis | | | Purpura | |
| Cutaneous lupus | | | Pyoderma Gangrenosum | |
| Cutaneous lymphadenoma | | | Pyogenic granuloma | |
| Cutaneous metastasis | | | | |
| Cutaneous myiasis | | | | |
| Cutaneous neurofibroma | | | | |
| Cutaneous neuroma | | | | |
| Cutaneous sarcoidosis | | | | |
| Cutaneous schistosomiasis | | | | |
| Cutaneous sporotrichosis | | | | |
| Cutis laxa | | | | |
| Cutis verticis gyrata | | | | |
| Cylindroma of skin | | | | |
| **Cyst** | | | | |

**Supplementary Table 7 | Hyperparameters for training the deep learning system.**

| | |
|---|---|
| Image augmentations | Image size: 459×459 pixels<br>Saturation delta: [0.5597, 1.2749]<br>Contrast delta: [0.9997, 1.7705]<br>Brightness max delta: 0.1148<br>Hue max delta: 0.0251<br>Rotation: [-150, 150] (degrees)<br>Flipping: horizontal, vertical |
| Bounding box augmentations | Minimum overlap with any pathologic region: 0.2<br>Aspect ratio: [0.9, 1.1]<br>Proportion over the original image: [0.05, 1.0] |
| Metadata augmentations | Dropout rate: 0.1 |
| Learning rate schedule (exponential decay schedule) | Base rate: 0.001<br>Decay rate: 0.99<br>Number of epochs per decay: 2.0 |
| Adam optimizer | Decay for the first moment estimates: 0.9<br>Decay for the second moment estimates: 0.999<br>Epsilon: 0.1 |
| Batch size | 8 |
| Regularization | Prelogits dropout rate: 0.2<br>Weight decay: 0.00004<br>Batch norm decay: 0.9997 |
| Loss function | Softmax cross-entropy with class-specific weights |
| Class weighting | Weight for each class is determined as: $1/c^{1-s}$<br>Where c is the class counts over the training set, and s is a smoothing factor of 0.7. |

**Supplementary Table 8 | Performance of the deep learning system (DLS) and different types of clinicians, on validation sets A and B.** This is similar to Extended Data Table 1, except performance was measured by the agreement of the top-1 and top-3 diagnoses with *any* of the panel of three dermatologists comprising the reference standard. In other words, whether the top k predictions of the DLS or clinician captures the primary diagnosis of any member of the panel. For agreement with a differential diagnosis based on the "votes" of the panel, see Extended Data Table 1. Numbers in square braces indicate 95% confidence intervals. Bold indicates the highest value within each column for validation set B.

| Dataset | "Grader" | Top-1 $Accuracy_{any}$ | Top-3 $Accuracy_{any}$ |
|---|---|---|---|
| Validation set A | DLS | 0.83 [0.82, 0.84] | 0.97 [0.97, 0.98] |
| Validation set B (enriched subset of set A) | DLS | **0.80 [0.77, 0.82]** | **0.96 [0.94, 0.97]** |
| | Derm | 0.77 [0.75, 0.79] | 0.86 [0.84, 0.88] |
| | PCP | 0.59 [0.56, 0.61] | 0.74 [0.72, 0.76] |
| | NP | 0.54 [0.52, 0.56] | 0.70 [0.68, 0.72] |

**Supplementary Table 9 | Performance of the deep learning system (DLS), stratified by self-reported demographic information (including age, sex, race and ethnicity), and Fitzpatrick skin type on validation set A.** Metrics used are identical to the ones in Supplementary Table 8. Numbers in square braces indicate 95% confidence intervals.

| Breakdown | Category | Top-1 Accuracy$_{any}$ | Top-3 Accuracy$_{any}$ |
|---|---|---|---|
| Age | [18, 30) (29.5%) | 0.86 [0.84, 0.88] | 0.98 [0.97, 0.99] |
| | [30, 40) (19.9%) | 0.82 [0.79, 0.85] | 0.97 [0.96, 0.98] |
| | [40, 50) (17.3%) | 0.82 [0.79, 0.85] | 0.98 [0.97, 0.99] |
| | [50, 60) (18.6%) | 0.83 [0.80, 0.85] | 0.97 [0.96, 0.98] |
| | [60, 90] (14.6%) | 0.78 [0.75, 0.82] | 0.97 [0.95, 0.98] |
| Sex | Female (63.1%) | 0.84 [0.83, 0.86] | 0.98 [0.97, 0.98] |
| | Male (36.9%) | 0.81 [0.79, 0.83] | 0.97 [0.96, 0.98] |
| Race and ethnicity | American Indian or Alaska Native (1.1%) | 0.79 [0.64, 0.90] | 0.98 [0.93, 1.00] |
| | Asian (12.6%) | 0.84 [0.81, 0.88] | 0.98 [0.97, 0.99] |
| | Black or African American (6.1%) | 0.86 [0.81, 0.90] | 0.97 [0.95, 0.99] |
| | Hispanic or Latino (43.4%) | 0.83 [0.81, 0.85] | 0.98 [0.97, 0.98] |
| | Native Hawaiin or Pacific Islander (1.6%) | 0.74 [0.62, 0.85] | 1.00 [1.00, 1.00] |
| | White (31.3%) | 0.82 [0.80, 0.85] | 0.97 [0.95, 0.98] |
| | Not specified (3.9%) | 0.84 [0.78, 0.90] | 1.00 [1.00, 1.00] |
| Fitzpatrick skin type | Type I (0.2%) | 0.78 [0.56, 1.00] | 0.78 [0.56, 1.00] |
| | Type II (10.2%) | 0.82 [0.78, 0.86] | 0.97 [0.95, 0.99] |
| | Type III (64.2%) | 0.83 [0.82, 0.85] | 0.98 [0.97, 0.98] |
| | Type IV (19.3%) | 0.83 [0.80, 0.85] | 0.97 [0.96, 0.98] |
| | Type V (2.7%) | 0.87 [0.80, 0.93] | 0.97 [0.93, 1.00] |
| | Type VI (0.0%) | 1.00* | 1.00* |
| | Unknown (3.4%) | 0.79 [0.72, 0.87] | 0.98 [0.95, 1.00] |

* : There was only 1 case labeled as Type VI, so confidence intervals were not meaningful.